# Walk Along: An Experiment on Controlling the Mobile Robot ‘Spot’ with Voice and Gestures

**April 2025**

R. Zhang*, J. van der Linden*, D. Dodou, H. Seyffert, Y. B. Eisma[#], J. C. F. de Winter[#,$]

*Faculty of Mechanical Engineering, Delft University of Technology, Delft, The Netherlands*

*Joint first authors
[#]Joint last authors
[$]Corresponding author: j.c.f.dewinter@tudelft.nl

**Abstract**
Robots are becoming more capable and can autonomously perform tasks such as navigating between locations. However, human oversight remains crucial. This study compared two touchless methods for directing mobile robots: voice control and gesture control, to investigate the efficiency of these methods and the preference of users. We tested these methods in two conditions: one in which participants remained stationary and one in which they walked freely alongside the robot. We hypothesized that walking alongside the robot would result in higher intuitiveness ratings and improved task performance, based on the idea that walking promotes spatial alignment and reduces the effort required for mental rotation. In a 2×2 within-subject design, 218 participants guided the quadruped robot Spot along a circuitous route with multiple 90° turns using rotate left, rotate right, and walk forward commands. After each trial, participants rated the intuitiveness of the command mapping, while post-experiment interviews were used to gather the participants’ preferences. Results showed that voice control combined with walking with Spot was the most favored and intuitive, whereas gesture control while standing caused confusion for left/right commands. Nevertheless, 29% of participants preferred gesture control, citing increased task engagement and visual congruence as reasons. An odometry-based analysis revealed that participants often followed behind Spot, particularly in the gesture control condition, when they were allowed to walk. In conclusion, voice control with walking produced the best outcomes. Improving physical ergonomics and adjusting gesture types could make gesture control more effective.

## Introduction

Robots have traditionally been viewed as devices designed to perform repetitive tasks, mainly in industrial settings and logistical operations. However, with the growing capabilities of AI, robots are taking on new roles. Modern robots can understand and adapt to their surroundings, paving the way for mobile robotics. This field extends beyond the familiar household cleaning robots and now also encompasses drones (Halder & Afsari, 2023; Roldán-Gómez et al., 2021), surveillance robots (Chen et al., 2021; Hafezi et al., 2024), underwater robots (Brantner & Khatib, 2021; Nauert & Kampmann, 2023), warehouse robots (Fottner et al., 2021; Jacob et al., 2023), agricultural robots (Benos et al., 2023; Gonzalez-de-Santos et al., 2020), and assistant robots (Hong et al., 2022; Mišeikis et al., 2020), among others.

The human-machine interface (HMI) plays a vital role in the control of mobile robots, as these robots are not yet capable of fully autonomous operation in open-ended environments (e.g., Endsley, 2017; Ezenkwu & Starkey, 2019; Hatanaka et al., 2023; Pianca & Santucci, 2023). Although mobile robots can execute certain maneuvers or subtasks, human supervision and interaction are still needed for overall task completion. Traditional HMIs for controlling mobile robots include joysticks (Dinh et al., 2017; Moniruzzaman et al., 2022), gamepads (Solanes et al., 2022; Wan et al., 2023), keyboards (Di Vincenzo et al., 2022), and graphical user interfaces (GUI) on tablets carried by human operators (Colceriu et al., 2023; Kaczmarek et al., 2021).

**Voice and gestures as natural interfaces**

While traditional HMIs have demonstrated their effectiveness over the years, they are not necessarily the most ideal solutions (as pointed out by Gorecky et al., 2014). Instead, an ideal user interface for mobile robots might mimic human-to-human or human-to-animal interactions (Krueger, 1993). Given that the human brain has evolved specialized areas for processing speech and body language, using such natural user interfaces could require less mental effort compared to interactions via handheld devices or other physical media. Hancock (1993), ahead of his time, proposed the ‘sheepdog metaphor’ for human-machine collaboration. This metaphor describes the relationship between a human (the shepherd) and a robot (the sheepdog). Just as a shepherd gives basic commands to a sheepdog, allowing it to manage subtasks based on its instincts and learned experience, a human operator can set high-level targets for a mobile robot, which then performs these tasks independently. The sheepdog metaphor illustrates that a limited set of commands can facilitate effective human-machine collaboration. More recently, and in accordance with the sheepdog metaphor, the concept of maneuver-based control has also been explored in automated driving (Detjen et al., 2020; Fink et al., 2023).

Possible candidates for interaction between humans and mobile robots include voice control (Li et al., 2023; Naeem et al., 2024) and control by means of mid-air gestures (Carfì & Mastrogiovanni, 2021; Coronado et al., 2017). Beyond technical issues regarding the detectability of the commands, such as voice control being less effective in noisy environments (Brunete et al., 2021; D’Attanasio et al., 2024) or gesture control being impractical when someone is wearing gloves or when otherwise constrained (Hatscher & Hansen, 2018; Sadhu et al., 2023), there are human factors to consider when implementing such HMIs. Voice commands are transient, while gestures can be sustained by the user for longer. It has been found that gesture control can generate a higher sense of embodiment than a traditional keyboard interface, although with a lower task performance (Di Vincenzo et al., 2022). Research has shown that when individuals were tasked with solving challenging spatial-visualization tasks (such as mentally rotating 3D objects), they spontaneously produced hand gestures and showed improved performance when doing so (Chu & Kita, 2011). These findings indicate that gestures can serve as a bridge between physical movement and abstract spatial reasoning, and suggest that gestures might represent an intuitive interface for controlling a robot’s movements and rotations. With speech interfaces, on the other hand, an intended movement of the robot must first be mentally translated into a verbal command. At the same time, Norman (2010) argued that gestural interfaces are not inherently intuitive or easy to learn.

**Existing applications of voice and gestures for mobile robot control**

Previous research on voice versus gesture control for operating mobile robots (Abich & Barber, 2017; Chivarov et al., 2019), in car cockpits (Detjen et al., 2019), or for controlling visual interfaces (Flick et al., 2021) shows that users generally prefer voice control over gesture control (but see Korkiakoski et al., 2024, for mixed findings). Differences in the robustness of detection and the simplicity or trainability of gestures (Nogales & Benalcázar, 2021; Zhou et al., 2023) may partly explain the relatively low ratings attributed to gesture control in some of the prior research. Factors such as the social acceptability of commands (e.g., speaking or gesturing in a shared space may confuse others; Rico & Brewster, 2010), competition with other tasks (e.g., drivers needing to keep their attention on the road and their hands on the wheel; Bilius & Vatavu, 2021), and recognition time delays (e.g., Halder et al., 2024) also play important roles.

As for quadruped mobile robots, previous studies investigated giving commands via gestures and/or body pose (Rippy et al., 2024; Sandberg, 2023; Shin et al., 2024; Steinke et al., 2023) or via voice, eye gaze, and head pose (Zhang et al., 2023). Despite several technical limitations needing resolution, these innovative methods of control showed promise. Other studies demonstrated the potential of quadruped mobile robots for guiding people with visual impairments (Cai et al., 2024; Chen et al., 2023; Doore et al., 2024; Due, 2023; Hata et al., 2024; Kim et al., 2025) or examined people's perceived safety as a function of the robot's walking style (Hashimoto et al., 2024). Of particular interest is a study by Chacón Quesada and Demiris (2024), which evaluated the effectiveness of an AR interface compared to a traditional handheld interface for controlling a quadruped robot. Participants completed navigation and manipulation tasks, including a condition that incorporated a cognitive offloading feature into the AR interface. In this context, cognitive offloading refers to allowing users to physically position themselves and use hand gestures, such as pointing to where the robot should go and using voice commands to direct its actions. The study found that with cognitive offloading, the AR-based interface yielded faster task completion times, reduced mental workload, and increased usability ratings.

For a complete comparison between voice and gesture controls in mobile robot operation, user orientation relative to the robot has to be considered. Research on stimulus-response compatibility shows that a difference in orientation between stimulus and response (requiring mental rotation) increases errors and information processing time (Fitts & Seeger, 1953; Shepard & Metzler, 1971).

Wickens and Prevett (1995) identified two HMI display perspectives: egocentric and exocentric. An egocentric display matches the user's viewpoint, since control inputs correspond with the vehicle's/robot's direction, as in first-person-view (FPV) displays used by expert drone racing pilots (Pfeiffer & Scaramuzza, 2021; Tezza et al., 2021). In contrast, an exocentric display, such as a top-down view or bird's-eye view, provides a detached perspective, which may increase the user's situational awareness but complicates control input generation (Smolyanskiy & Gonzalez-Franco, 2017). In practice, the user's mobility and positioning relative to the robot will result in different perspectives: egocentric for users closely aligned with the robot's orientation and exocentric for those observing from a distant or static perspective, both of which are likely to influence performance in control tasks.

To the best of our knowledge, no comparative study has investigated voice and gesture interfaces for mobile robots while accounting for differences in users' mobility (e.g., walking vs. standing). Most existing research that used voice or gestures to control a mobile robot combined these modes or lacked a direct comparative analysis (Chacón Quesada & Demiris, 2024; Doore et al., 2024; Du et al., 2021; Fardana et al., 2013; Hata et al., 2024). In a previous study involving older adults and individuals with disabilities controlling a mobile robot, gesture control was found to be less preferred and more fatiguing than voice or joystick control (Chivarov et al., 2019). Abich and Barber (2017) compared speech and sign-language-based gestures for robot navigation and investigation tasks. However, because the study replaced each spoken command with an artificial one-to-one gesture mapping based on sign language (unlike natural/spontaneous gestures), its comparison between voice and gesture control may be hard to interpret.

Existing gesture interfaces for mobile robot control have relied on wearable gloves (Abich & Barber, 2017) or an inertial measurement unit (IMU) to track hand movements (Kandalaft et al., 2018; Meghana et al., 2020) and non-head-mounted cameras to track body posture (Fardana et al., 2013; Sandberg, 2023; Shin et al., 2024), hand position (Chivarov et al., 2019) or finger counts (Vanamala et al., 2022), which limited the users' mobility and might not be effective in mobile scenarios. We chose to use an augmented reality (AR) headset that tracks the user's inputs well in both mobile and stationary conditions. The vision-based gesture recognition we implemented does not require the participants to wear sensors or stay in front of cameras or microphones, leading to natural hand and body movements. Furthermore, we have published our code as open-source, allowing others to reproduce our setup and build upon it.

**A comparative study of voice and gesture control in standing and walking conditions**

We conducted an experiment using the AR device Microsoft HoloLens 2 to compare voice and gesture control for operating Spot, a popular quadruped robot developed by Boston Dynamics. In the current study, participants navigated the robot along a trajectory with multiple 90° turns using each control method twice: once while standing still at the starting point and once while walking alongside the robot. When standing still, participants had a consistent view of the trajectory but experienced stimulus-response incompatibility, especially when the robot was oriented toward or approaching them. For instance, a *Rotate Right* command in that case would make the robot rotate left from the participants' perspective. When walking, participants could position themselves behind or next to the robot to reduce this incompatibility.

We examined the effects of these two independent variables (1. control method, 2. participant mobility) on task performance and self-reported experience. We differentiated between the detectability of the commands (i.e., a technical issue regarding speech recognition and computer vision) and the intuitiveness of command-to-robot-movement (a human factors issue). We hypothesized that walking with the robot would result in better task performance and higher intuitiveness ratings than controlling the robot while standing still.

Additionally, we expected that gesture control while standing still would be particularly unintuitive due to a spatial mismatch between the participant's hand gestures and the robot's movement. In contrast, voice control, which involves verbal and auditory processing, conflicts less with visual tasks according to Wickens's multiple resource theory (Wickens, 2002, 2008). This theory posits that humans have distinct cognitive resources (visual, auditory, spatial, and verbal), and tasks using the same resource type interfere more with each other. Therefore, voice control should cause less cognitive conflict during moments of incompatibility compared to gesture control, which relies heavily on visual-spatial resources.

## Method

### Participants

A total of 218 participants took part in this experiment. Of these, 216 were students enrolled in an MSc course on Human-Robot Interaction offered by the Faculty of Mechanical Engineering at TU Delft. The remaining two participants were staff members from the same faculty. The course was part of the MSc Robotics program and the Biomechanical Design track within the MSc Mechanical Engineering program at TU Delft. Consequently, it can be assumed that the students had a strong interest in, or basic to advanced knowledge of, technical and robotic systems. However, the HoloLens 2 in combination with the Spot robot was introduced to the students for the first time, meaning they were unfamiliar with the experimental facility and task. The experiment was conducted from December 4, 2023, to January 19, 2024.

In addition to these 218 participants, 3 participants could not complete the experiment because their control inputs were not recognized, presumably due to an issue with the wireless connection during the first two days of the experiment, and were excluded from consideration. Because our interest was in making a complete comparison between gesture control and voice control, we did not exclude participants whose hand gestures or voice commands were less well recognized during the experiment.

Participants' ages, collected via a post-experiment questionnaire, ranged from 21 to 30 years, with a mean age of 23.5 years (*SD* = 1.64). Among the participants, 194 were right-handed, 18 were left-handed, and 6 were mixed-handed. Gender distribution included 160 males, 55 females, and 3 individuals who preferred not to disclose their gender. Out of the participants, 121 reported never wearing any vision aids, 49 wore glasses during the experiment, 36 wore contact lenses, and 12 typically wore glasses or contact lenses but did not wear them during the experiment. In response to a question: "*Before the experiment, did you ever wear a HoloLens or similar augmented-reality device?*", 155 participants answered *no*, and 63 participants answered *yes*.

The experiment was approved by the Delft Human Research Ethics Committee (HREC), approval no. 3502, with each participant providing written informed consent before the experiment.

**Hardware and software**

The experiment setup included the quadruped robot Spot Explorer (Boston Dynamics, 2020), measuring 1100 mm in length, 500 mm in width, and 840 mm in standing height. The setup included a HoloLens 2 (Microsoft, 2024a), a device used in combination with Spot in earlier studies (Chacón Quesada & Demiris, 2024; Chen et al., 2024; Rippy et al., 2024; Soler et al., 2024; Steinke et al., 2023; Vom Hofe et al., 2023; Zhang et al., 2023). Additionally, a Windows PC equipped with an Intel i7-8700K processor and an NVIDIA RTX 2080 graphics card was used. These devices were connected to a TP-Link AX3000 Gigabit Wi-Fi 6 Router (TP-Link, 2019), with the PC connected via a network cable and the HoloLens 2 and Spot connected via a 2.4 GHz Wi-Fi network.

The HoloLens 2 was used to capture the participant's control inputs. It ran an application developed in Unity v2021.3.28f1 (Unity, 2021). The application was created using the Mixed Reality Toolkit (MRTK) v2.8 (Microsoft, 2022a) for interaction design and UI components. The OpenXR Plugin was used to ensure compatibility with the HoloLens platform. Communication between the HoloLens 2 and the PC took place via the Robot Operating System (ROS) (Quigley et al., 2009) and ROS-TCP-Connector package (Unity, 2022). Spot was controlled from the PC using the Python interface of Spot-SDK 3.3.2 (Boston Dynamics, 2023). An overview of the hardware and software infrastructure is provided in Figure 1.

**Task**

The participants were tasked to control the robot either using their voice or gestures to direct Spot along a trajectory on the floor as quickly as possible. The trajectory contained 14 virtual nodes, including its start and end nodes (see Figures 2 & 3). Participants needed to use three maneuvers/commands: *Walk Forward*, *Rotate Left*, and *Rotate Right*, to complete the trajectory. The *Walk Forward* command moved Spot 1 m forward in its Crawl locomotion mode, in which three feet touch the ground at all times. The default velocity limit for crawling was used, resulting in an execution duration of 3 s for each command. Crawl mode was selected because Spot walks more slowly in this mode compared to the Walk (Trot) mode. It was reasoned that the faster Walk mode could appear intimidating and reduce the participants' willingness to walk alongside the robot. The *Rotate Left* and *Rotate Right* commands rotated Spot 90° to its left and right, respectively. The task was completed with a minimum number of 23 commands, including 13

*Walk Forward*, 6 *Rotate Left*, and 4 *Rotate Right* commands, resulting in 23 checkpoints of the robot pose.

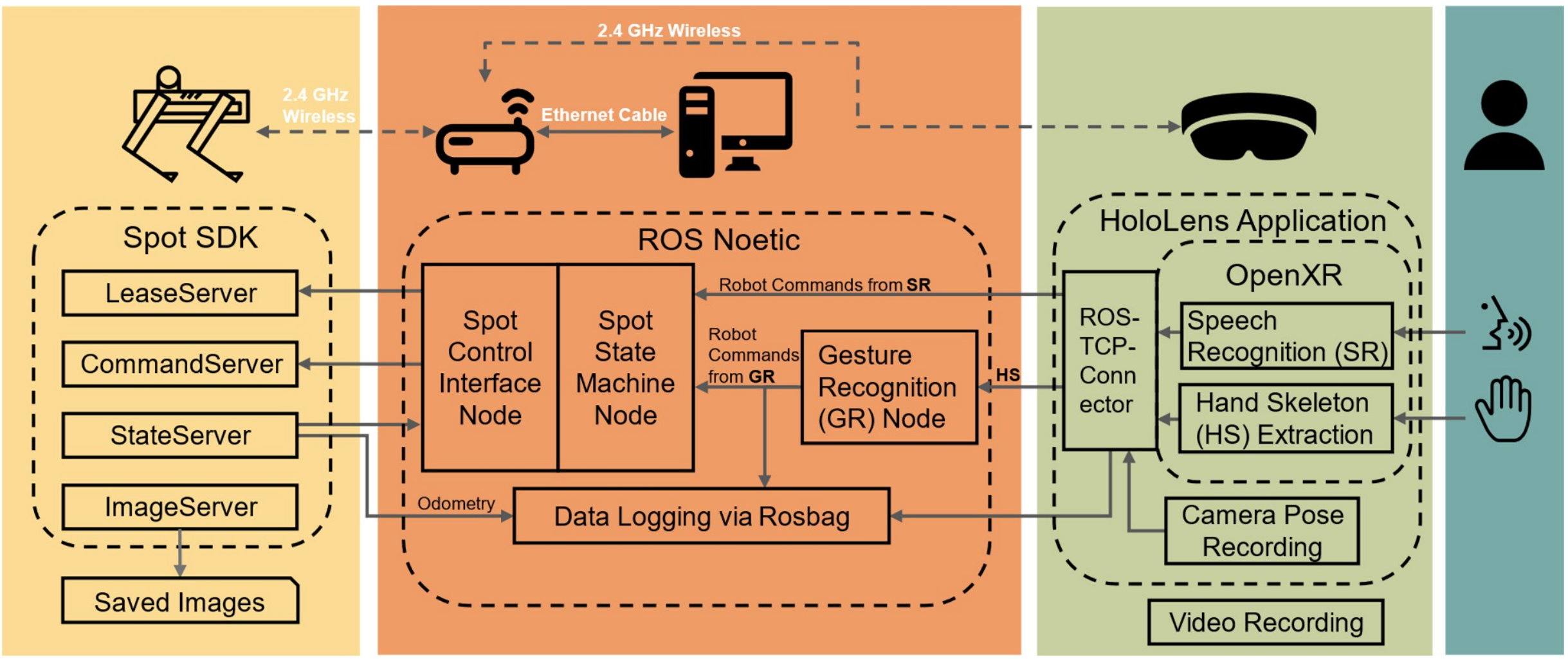

*Figure 1.* The hardware and software structure of the setup. The HoloLens 2 captures human user inputs (i.e., speech and hand skeleton data) and processes them via a HoloLens application, which was developed in Unity (Unity, 2021) with the OpenXR Plugin for Unity, managed by the Mixed Reality Toolkit 2 (MRTK) (Microsoft, 2022a). The application also records the 6D pose of the central front camera of the HoloLens 2. The processed user inputs from the plugin and the camera's 6D pose are recorded via Rosbag. Additionally, a corresponding video of the user's view is recorded by the HoloLens 2. Robot commands are recognized either from speech by the plugin or from gestures by a ROS node with the hand skeleton data from the application. A state machine manages the actions of the Spot robot and controls it via the Spot Software Development Kit (SDK).

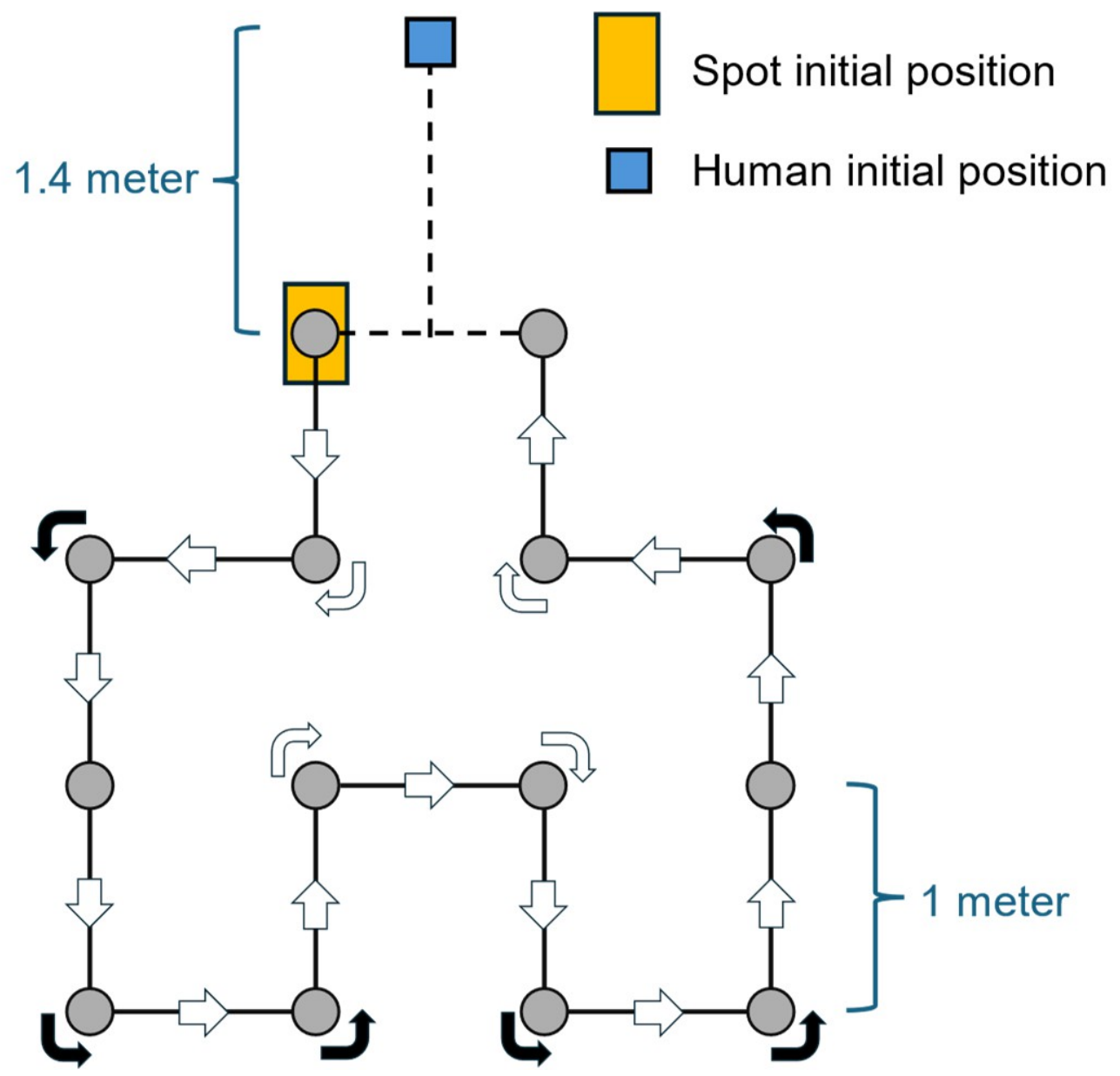

*Figure 2.* The target trajectory of the Spot robot. The 14 nodes are indicated by means of gray circles. The *Walk Forward*, *Rotate Left*, and *Rotate Right* commands are represented by straight arrows, black curved arrows, and white curved arrows, respectively.

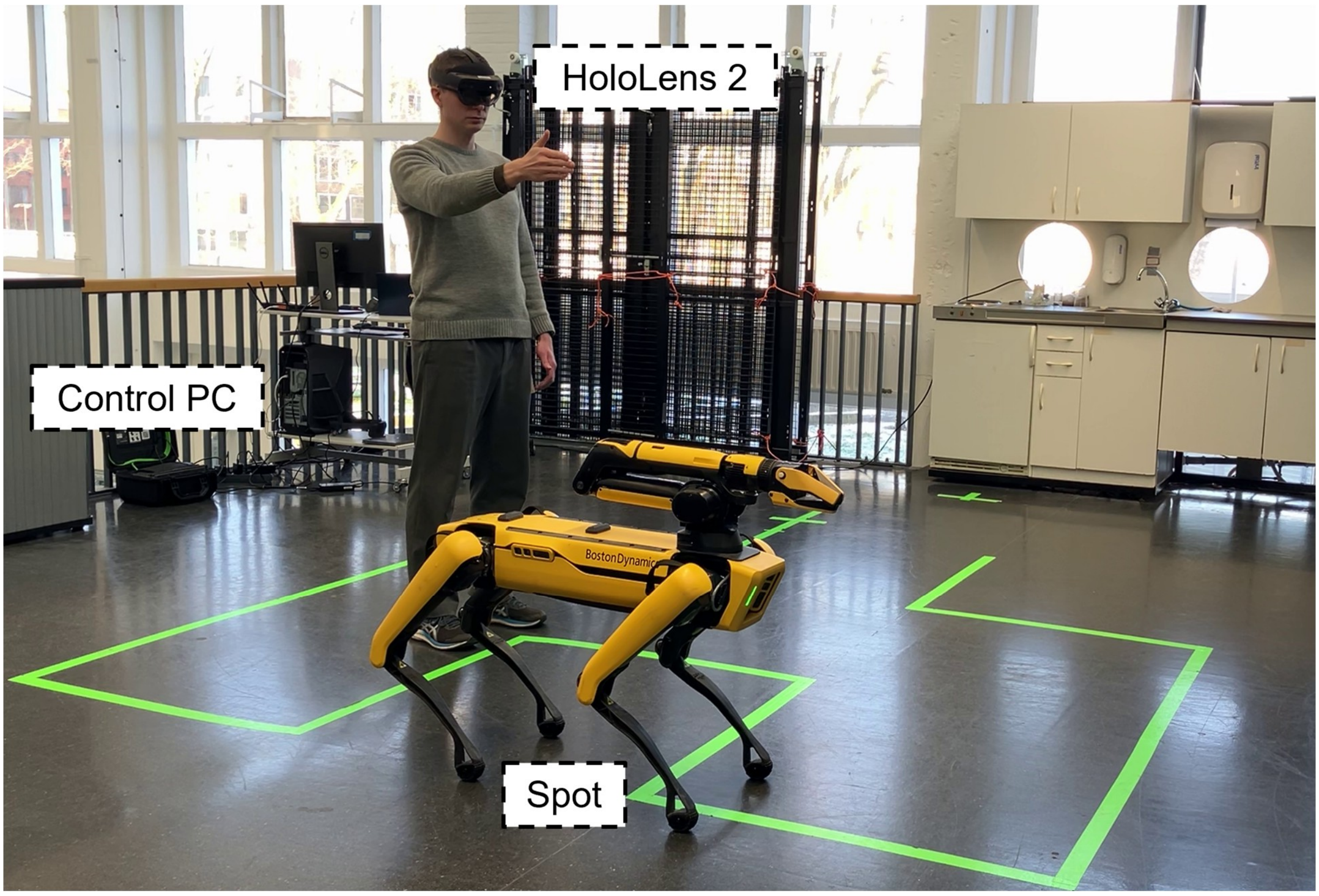

*Figure 3*. Overview of the experimental setting. A participant wearing the HoloLens 2 controlled Spot using the gesture interface while walking alongside it.

**Independent variables**

There were two independent variables, each consisting of two levels, and both were manipulated within subjects:

1. *Participant's mobility.* Participants were instructed to either remain in their initial position or walk with the robot.
2. *Control method.* Participants used either voice or gestures to control Spot. In the voice control conditions, participants vocalized one robot command at a time and then waited for the robot to execute it. Similarly, in the gesture control conditions, participants performed a gesture, held it until it was recognized, and then waited for the robot to execute the command.

These two independent variables resulted in the four experimental conditions shown in Table 1. The experimental conditions were fully counterbalanced, with the 24 possible sequences of the four experimental conditions repeating every 24 participants.

**Speech recognition**

The HoloLens features speech recognition that can be customized within Unity using the MRTK (Microsoft, 2022b). In this experiment, three voice commands were programmed to be recognized in order to control the robot, namely *Walk Forward*, *Rotate Left*, and *Rotate Right*. A pilot study showed that participants took about 1 s to utter a command, with an additional 1.2 s required for the system to process it. Thus, the entire process from speaking to recognition took approximately 2.2 s.

Table 1

*The four conditions tested in the experiment. Participants were instructed to use either voice (V) or gesture (G) to control the robot while walking (W) with it or standing (S) at the initial position.*

| | | **Participant's mobility** | |
|---|---|---|---|
| | | *Walking* | *Standing* |
| **Control method** | *Voice* | Voice control–Walking (VW) | Voice control–Standing (VS) |
| | *Gesture* | Gesture control–Walking (GW) | Gesture control–Standing (GS) |

**Gesture recognition**

A custom gesture recognition pipeline was implemented based on the MRTK of the HoloLens (Figure 4). The HoloLens provides the 3D coordinates of detected hand skeletons in 26 joints (25 joints from the hand skeleton and 1 joint of the palm) with respect to the HoloLens frame at a frequency of approximately 30 Hz (Microsoft, 2020, 2021). The data of the hand skeletons from the HoloLens were forwarded to the PC and classified into one of four gestures using a Support Vector Machine (SVM) trained by previously collected gesture data from the HoloLens (see Appendix A). Three gestures were used for robot commands (see Figure 5F). One gesture (fist) acted as a neutral gesture, similar to silent periods between voice commands.

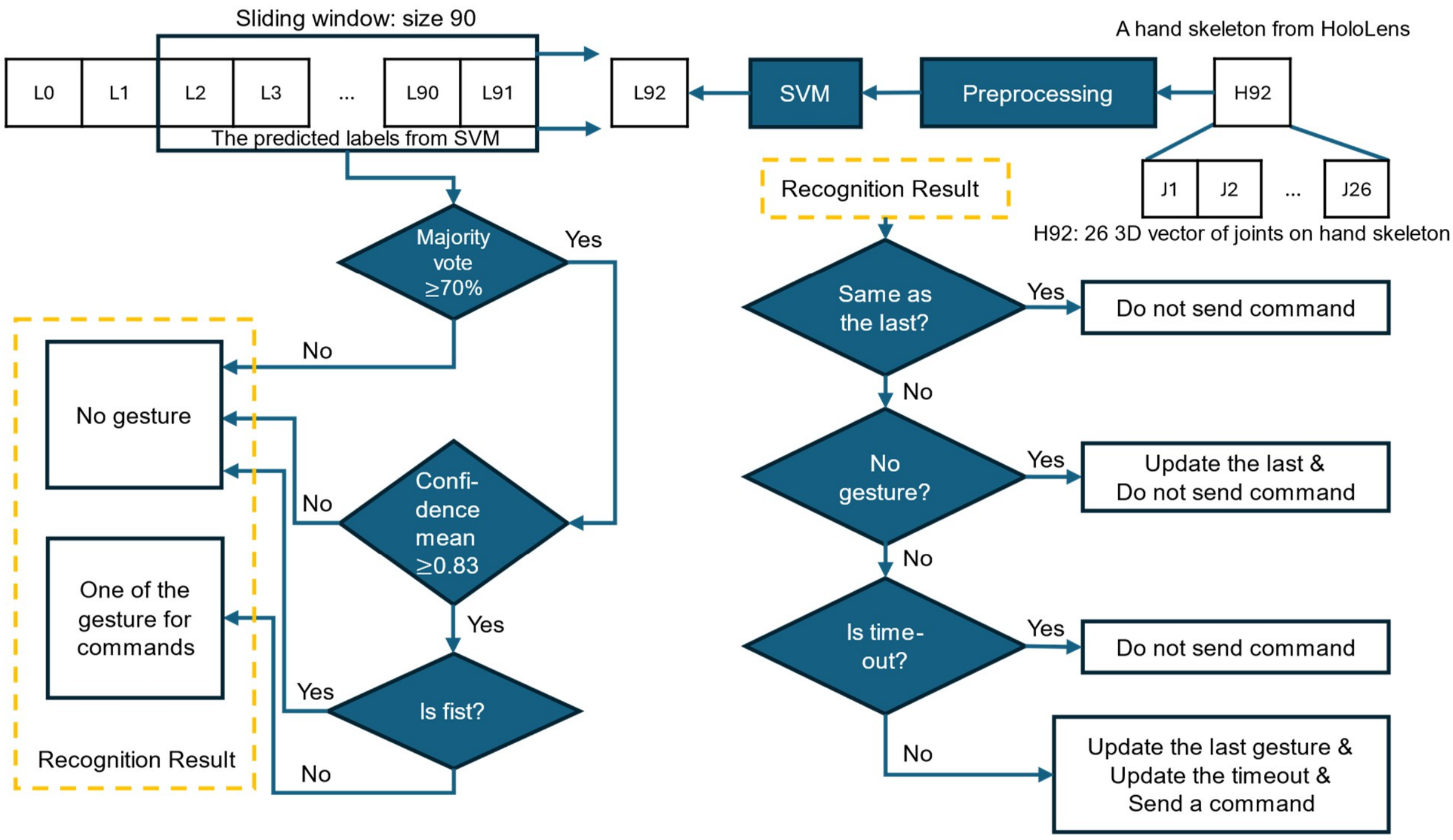

*Figure 4*. Gesture recognition process. Each hand skeleton frame detected by the HoloLens 2 was streamed to the PC as 3D coordinates of 26 joints and filled into a sliding window. A Support Vector Machine (SVM) gesture classifier predicted gestures in the latest 90 frames (frame rate approximately 30 Hz). These predictions were processed by a majority voting and a confidence check to determine the current hand gesture, which was sent as a valid gesture command only if it was a new (other than the fist/neutral) gesture and the timeout was not activated.

Hand skeleton data from the HoloLens were streamed to the SVM gesture classifier, and a sliding window system was applied to the predicted labels from SVM. The sliding window buffered the last 90 predictions of the incoming hand skeletons. This buffer remained unchanged if the HoloLens did not capture any hand skeleton data. A gesture was recognized when at least 70% (i.e., 63 or more samples) of the predicted labels within the current buffer were in the same class, and the mean classification confidence level of all 90 samples was above 0.83. Given that the recording frequency was approximately 30 Hz, the minimum time required for recognition was approximately 2.1 s. If the current gesture was the neutral gesture or the same as the last recognized gesture, no command was sent to the robot. After a gesture command was successfully sent to the robot, a 3-s timeout was activated during which no further recognized commands would be transmitted.

**Heads-up display information**

Once the participant put on the HoloLens, a series of visual AR elements appeared to guide the participant through each phase of the experiment (Figure 5). After the participant completed the built-in HoloLens instructions and performed the eye-tracking calibration (Figure 5A), our custom application was initialized. Initially, the first guidance panel and two buttons to start the experiment or to exit the application were displayed (Figure 5B). Upon pressing the start button (Figure 5C), the initial panel was replaced by the second panel and two switches, corresponding to gesture and voice control (Figure 5D). This guidance panel instructed the participant to follow the demonstration given by the experimenter (Figure 5E) and to toggle the appropriate switch to proceed with the upcoming condition (Figure 5F).

Once a switch was toggled, the second guidance panel faded out (Figure 5G). A command panel with the three available commands remained available (gesture commands: Figure 5F; voice commands: Figure 6). The command panel served as a memory aid for the participant, so that they could always refer to the three possible commands. Upon recognition of a voice or gesture command, a popup message (*Walk Forward*, *Rotate Right*, or *Rotate Left*) was displayed for 1.5 s (Figure 5G & 6). Additionally, the HoloLens emitted a beep tone approximately 3 s after the command was recognized, to indicate that the participant could issue the next command. The execution of the robot's movement lasted 3 s, commencing with the appearance of the popup message and concluding approximately at the same time as the beep.

At the end of each trial, participants deactivated the gesture or voice control (Figure 5I), which removed the command panel and activated a questionnaire panel along with a confirmation button (Figure 5J). Pressing the confirmation button (Figure 5K) closed the questionnaire panel and restored the second instruction panel, preparing for the next condition (Figure 5L).

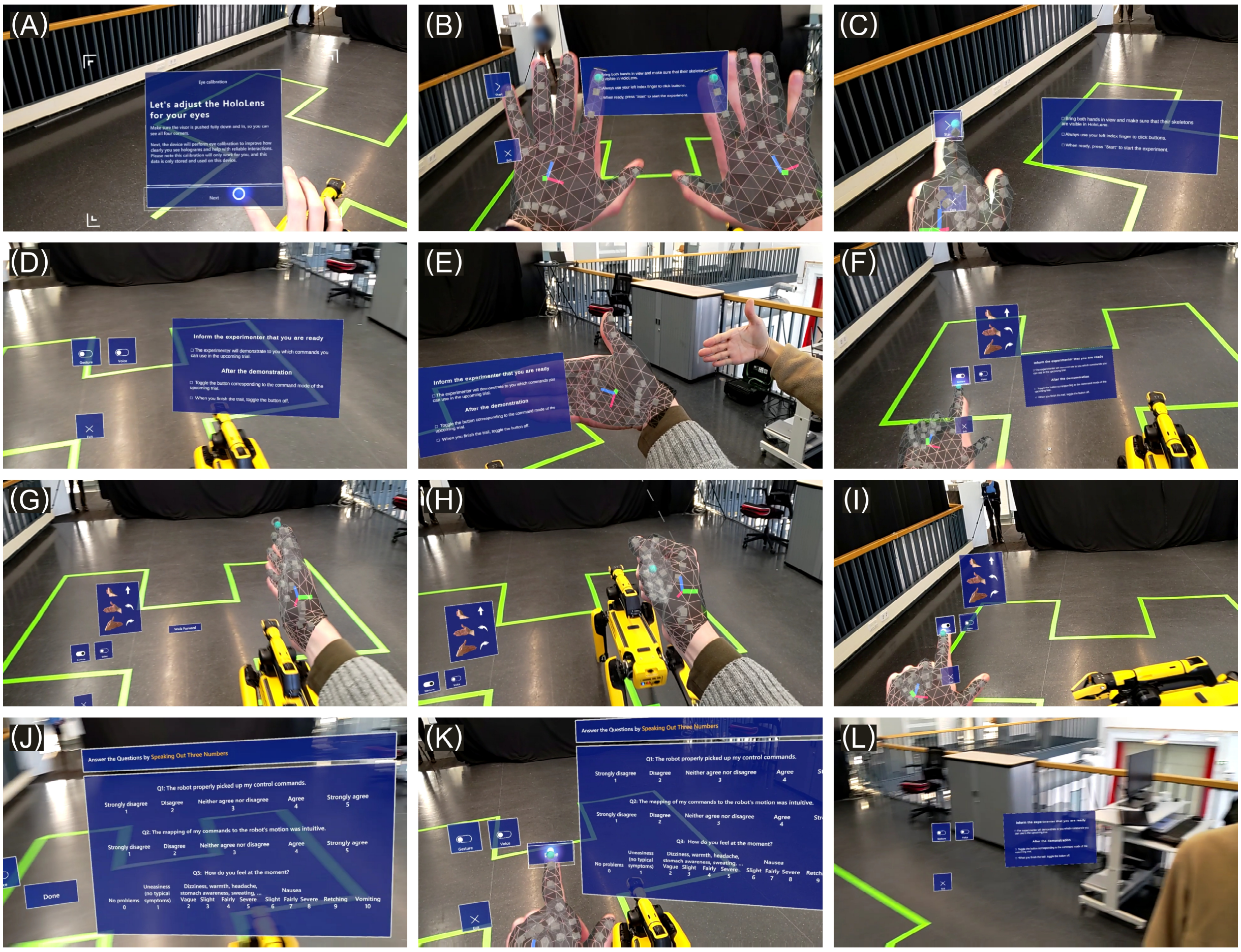

*Figure 5.* HoloLens interfaces during various phases of the experiment under the Gesture control–Standing (*GS*) condition. (A) The participant pressed the eye-tracking calibration button. (B) After the calibration, the participant saw the initial guidance panel with general instructions, along with “Start” and “Exit” buttons. (C) The participant pressed “Start”. (D) A second guidance panel, informing the participant about a demonstration from the experimenter and how to start a trial, was visible along with two switches for gesture and voice control. (E) The participant was mimicking the experimenter’s demonstration of commands. (F) Toggling gesture control made a command panel visible, while the second guidance panel was fading out. (G) A popup message confirmed a successfully recognized gesture command. (H) The participant continued the trial. (I) The participant toggled off the gesture control upon completion of the trial. (J) A post-trial questionnaire panel and a “Done” button were shown. (K) The participant pressed “Done” to finalize the questionnaire. (L) The questionnaire panel disappeared, and the second instruction panel was resumed.

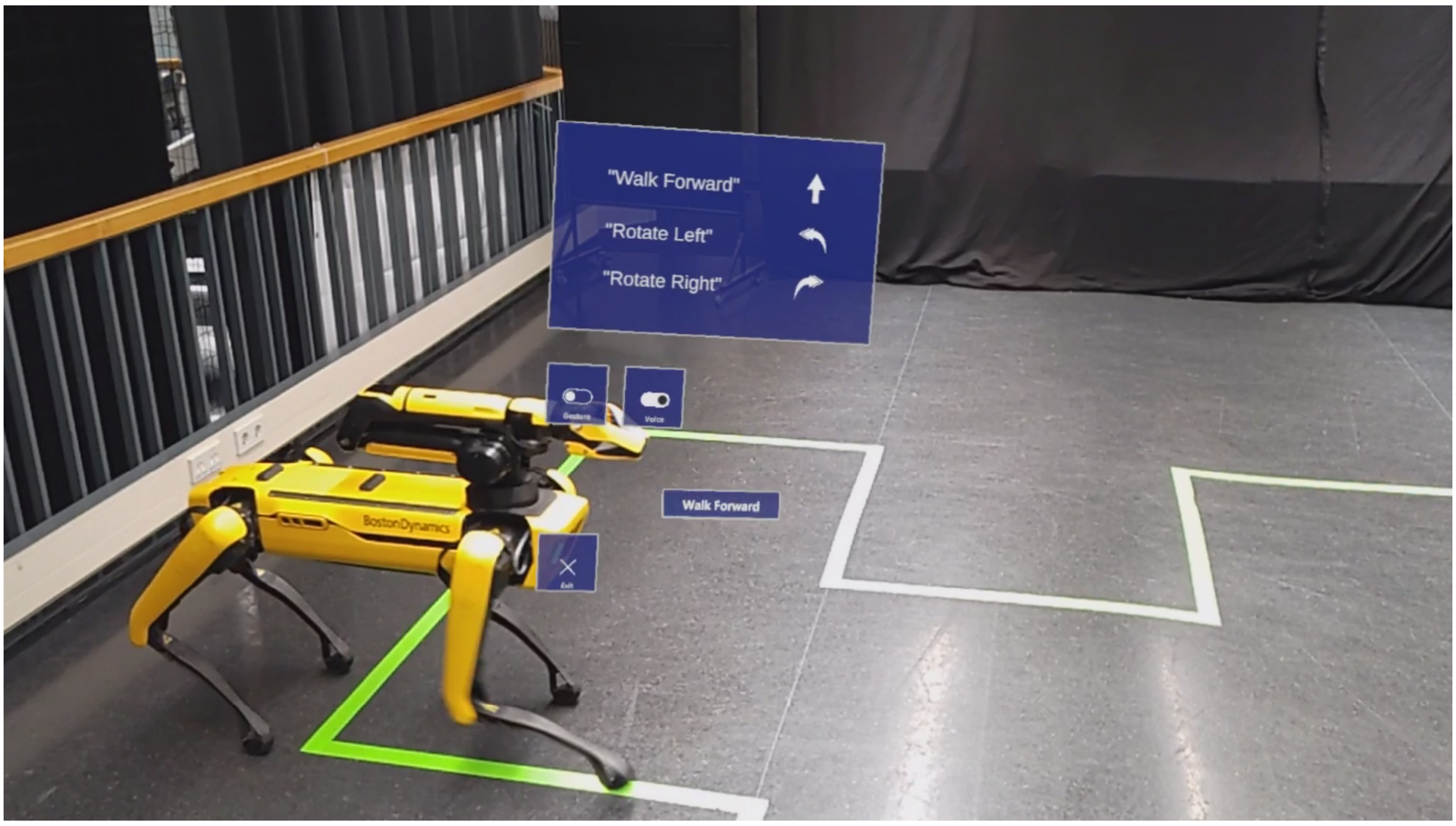

*Figure 6.* HoloLens 2 view for the Voice control–Standing (*VS*) condition. The command panel can be seen, as well as the popup window *Walk Forward* that indicates the latest recognized command.

**Dependent measures**

The collected data were used to examine participants' performance and behaviors. These measures were determined based on the stored command input data for Spot, the odometry of Spot (recorded at a frequency of approximately 30 Hz), and the data from the participants' HoloLens camera position and orientation (recorded at a frequency of approximately 25 Hz). For odometry, we used the "odom" frame, an inertial frame that provides an estimate of Spot's location in the world based on Spot's kinematics (the recorded data also contained a "vision" frame, which we did not use).

In addition to evaluating participants' performance and behaviors, we gathered their self-reported opinions on two key aspects: (1) Spot's command detection performance, i.e., whether participants felt the robot accurately interpreted their instructions (a technical issue related to speech or gesture recognition), and (2) the intuitiveness of the mapping between the participants' commands and Spot's movements (which reflects a psychological rather than technical issue). These two items were chosen over existing usability scales, which were considered too generic for this context.

Below, we report all the dependent measures used:

A. *Median inter-command time*. This measure represents the median time between sequential commands given to the robot. The median was chosen for its robustness against outliers of individual commands, such as occasional long responses due to interruptions or slow command detection.
   Shorter inter-command times are indicative of better human-robot interaction. To provide a benchmark, the minimum observed median inter-command time recorded among all 218 participants, including walking and rotating, was 3.3 s for the voice control conditions and 3.9 s for the gesture control conditions. These times represent performances achieved when participants issued commands in an anticipatory manner, i.e., before the robot completed its current maneuver.

B. *Number of commands*. This metric indicates the total number of robot commands issued in one trial to complete the trajectory. The minimum number of commands required was 23, as shown in Figure 2. Participants needed additional commands to correct any mistakes made during the trial. The more commands needed to be given, the worse the performance of the human-robot system.
C. *Total distance walked*. This was calculated from the *x* and *y* position coordinates of the HoloLens camera. A moving median filter with a time interval of 1 s was applied to the coordinates to remove the effect of high-frequency noise. The total distance covered by the participant is not a performance measure but an index of the participant's style of behavior. For the standing conditions, a distance close to 0 m is naturally expected. For the walking conditions, it is relevant to assess how much participants walked as well as their orientation relative to the robot (as expressed using Measure F).
D. *Command detection*, from 1 (Strongly disagree) to 5 (Strongly agree). Response to the question: "*The robot properly picked up my control commands*". This measure was based on a post-trial questionnaire (see Figure 7).
E. *Mapping intuitiveness*, from 1 (Strongly disagree) to 5 (Strongly agree). Response to the question: "*The mapping of my commands to the robot's motion was intuitive*". This measure was also based on the post-trial questionnaire.
F. *Participant-Spot alignment percentage.* To determine the extent to which participants oriented themselves in the same direction as Spot, we calculated the difference in bearing angle between the HoloLens camera and Spot. We defined 0° as when the participant and Spot were oriented in the same direction, 90° as Spot turned 90° to the right relative to the participant, 180° as Spot was oriented facing the participant, and 270° as Spot turned 90° to the left relative to the participant. We then calculated the percentage of time during the trial that this angle was between -45° (i.e., 315°) and +45°, as an index of the portion of time the human and robot were oriented in approximately the same direction. Note that participants had little control over this angle in the *VS* and *GS* conditions but had full control in the *VW* and *GW* conditions.

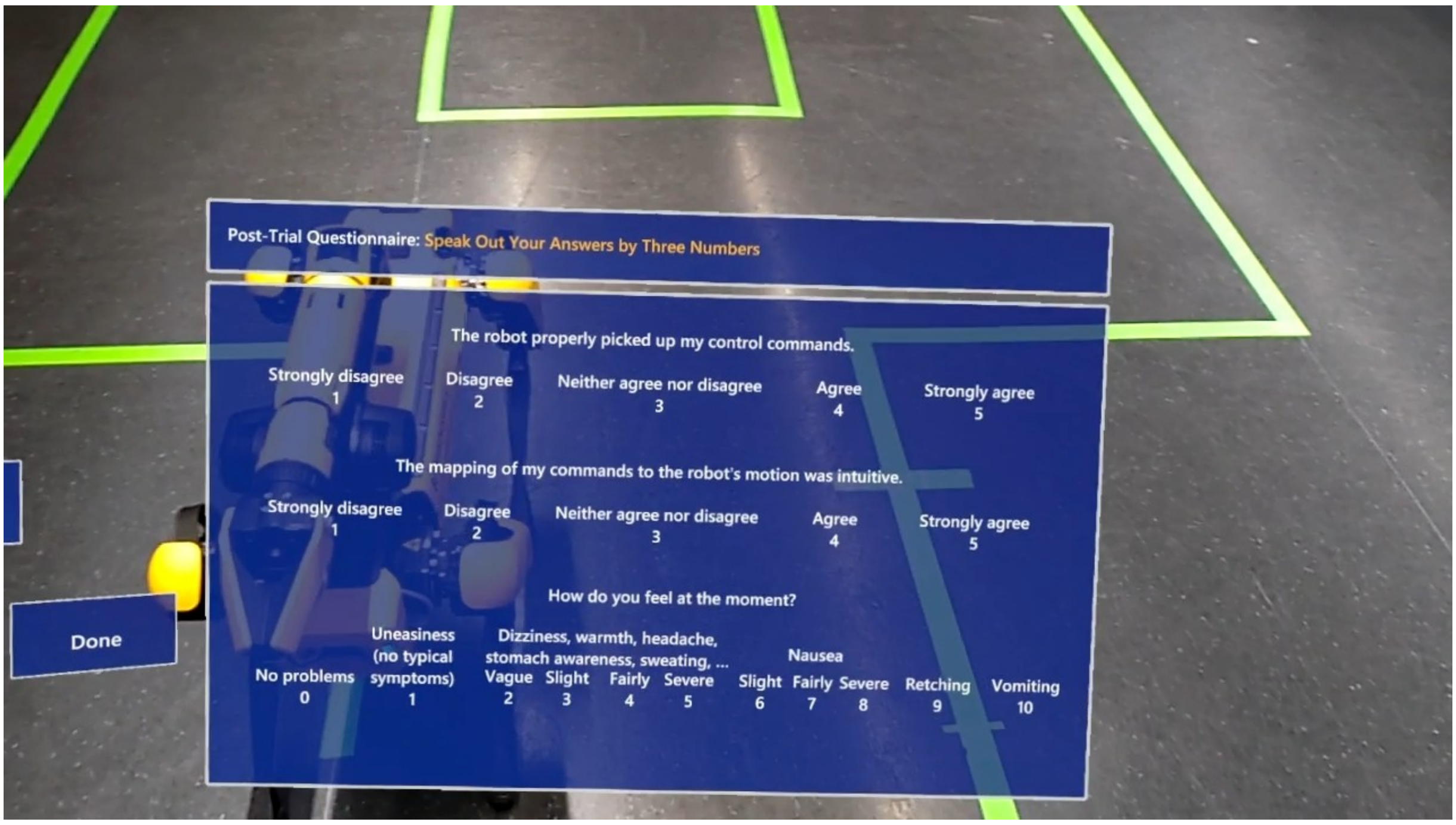

*Figure 7.* Post-trial questionnaire projected as an augmented reality overlay.

In addition to the above measures, we conducted a more specific performance analysis of the command inputs made by the participants. For each of the 23 checkpoints, we tabulated whether the participant gave an incorrect input, where we distinguished between duplicate commands, left/right mistakes, and other types of mistakes. Further details on how these errors were counted can be found in Appendix B.

G. *Number of duplicate command mistakes.* These are occasions where a participant mistakenly repeated the same command at a single checkpoint.
H. *Number of left/right mistakes.* These are occasions where, at a checkpoint, the participant should have provided a *Rotate Left* command but provided a *Rotate Right* command instead, or vice versa.
I. *Number of other command mistakes.* This category consists of cases in which a *Walk Forward* command was provided where *Rotate Left* or *Rotate Right* was expected, and cases in which *Rotate Left* or *Rotate Right* commands were provided where *Walk Forward* was expected.

Finally, a brief interview was conducted at the end of the experiment, where participants were asked to identify their most and least favorite conditions of the four tested and to explain why. The participants' answers were manually extracted from the HoloLens audio recording and counted. Subsequently, we counted the number of participants per condition and determined the corresponding 95% confidence intervals for the percentage of participants per category, as calculated based on a binomial distribution.

Previous research indicates that large language models (LLMs), provided that an appropriate prompting strategy is used, are capable of content analysis or summarizing interview transcripts (e.g., Friedman et al., 2024; Tabone & De Winter, 2023). Therefore, we performed an LLM-based analysis of the transcripts. Specifically, we transcribed all interviews using OpenAI's Whisper Large-V3 (Radford et al., 2023). Some interviews for which transcription failed were transcribed manually. We then used OpenAI's ChatGPT web interface (selected model: *o1 pro mode*; date:

16 December 2024; OpenAI, 2024) to ask a number of questions to gain further insight into the participants' preferences. We chose this model because previous evaluations suggested that the recent class of reasoning LLMs performs well in content analysis, i.e., tabulating after reflecting on texts (De Winter et al., 2025). One example prompt we used was: "*What are the five main reasons participants preferred voice control based on the transcripts? Provide concise one-liner statements (max 140 characters) summarizing the reasons, avoiding direct quotes.*", followed by all the transcripts.

Additionally, to evaluate the validity of the output generated by *ChatGPT o1 pro mode*, we prompted the model to examine whether it could replicate the participant counts based on the manual analysis of the participants' most favorite condition.

**Experimental procedure**

Upon arrival, participants were provided with a consent form. Participants also received a printout with the task instructions, stating:

*You will be asked to control the robot using four different control methods:*

- *Voice commands while standing*
- *Gesture commands while standing*
- *Voice commands while walking with the robot*
- *Gesture commands while walking with the robot*

*Your task is to instruct the robot to follow a designated trajectory and reach the end point as fast as possible.*

After signing the consent form, participants wore the HoloLens and performed the eye gaze calibration provided by the HoloLens (see Figure 5A). Next, the experimenter demonstrated the voice or gesture commands that could be used in the upcoming trial (see Figure 5E). In the voice control conditions, the three voice commands were given verbally as the demonstration. In the gesture control conditions, the experimenter used the right hand to demonstrate the three gestures corresponding to the robot commands and a neutral gesture, and instructed participants to use the right hand for gesture commands as well and to return to the neutral gesture between commands. Participants were informed that, during the trial, a panel (see Figures 5F & 6) would display the three voice or gesture commands. Depending on the experimental condition, the experimenter asked the participant to either stay in the initial position during the trial or walk with Spot in any way they preferred. Before the first trial, the experimenter explained that after a command was given, a small overlay would confirm the receipt of this command by the software (see Figures 5G & 6). It was also stated that a beep tone, occurring a few seconds later, would indicate when the next command could be given.

After the demonstration, the participants toggled the virtual button ("Voice" or "Gesture") to activate the corresponding control method and start the trial, marked by a beep tone. If the participant remained stationary during a walking trial, the experimenter reminded them to walk. During the trials, if a wrong command was given to the robot, the experimenter asked the participant to use commands to reposition Spot back to the last correct robot pose before the mistake. After completing a trial, the participants were asked to toggle the button off so that the post-trial questionnaire was displayed in the HoloLens (see Figure 7).

One trial was conducted in each condition for all participants. After completing all four conditions, the participants were briefly interviewed about which of the four conditions they favored most and least, and then asked to complete a post-experiment questionnaire using the Qualtrics platform (Qualtrics, 2024) on a laptop, to collect general participant information.

**Statistical analyses**

For each dependent measure, we calculated the mean and the 95% confidence interval for the mean, assuming a normal distribution. These means and confidence intervals were plotted in a bar plot with error bars. We also performed three paired-samples *t*-tests:

1) Voice control–Walking (*VW*) vs. Voice control–Standing (*VS*),
2) Gesture control–Walking (*GW*) vs. Gesture control–Standing (*GS*), and
3) Voice control–Walking & Standing (*VW&S*) vs. Gesture control–Walking & Standing (*GW&S*).

For the third comparison, the values for the standing and walking conditions were averaged per participant. To account for multiple comparisons, we reduced the critical alpha value to 0.05/3 ≈ 0.0167. Cohen's *d* was used to measure the effect size between the two conditions. The paired-samples *t*-test is based on the assumption that the differences between the paired observations should be approximately normally distributed. The use of non-normally distributed data or Likert-scale questionnaire items does not necessarily render the *t*-test unsuitable or require switching to non-parametric tests like the Wilcoxon signed-rank test. The suitability of the *t*-test depends on the degree to which assumptions are violated, and the associated Type I and Type II error probabilities. Research suggests that *t*-tests are generally appropriate for five-point Likert items (De Winter & Dodou, 2010).

In Appendix C, we substantiate the use of paired-samples *t*-tests by re-performing the statistical analysis after a rank transformation (Conover & Iman, 1981; Iman et al., 1984). Overall, the resulting *p*-values and corresponding Cohen's *d* effect sizes correspond to those obtained through the original paired-samples *t*-test. Since a paired-samples *t*-test is simpler to report and the associated means with confidence intervals are easier to visualize, we chose to use this method. Additionally, Appendix D includes boxplots of several dependent variables to increase transparency regarding the data distribution.

Regarding the angular difference between Spot and the participant, we generated a polar density plot to create a more complete picture of how the participant and Spot were oriented relative to each other overall. This density plot was calculated per participant per condition and then averaged over all participants.

## Results

**Trial completion and missing data**

All 218 participants each completed four trials, resulting in post-trial questionnaire data for all 872 trials. A recording of the post-experiment interview was unavailable for 1 out of 218 participants, due to a HoloLens failure. Regarding the commands, data for one trial in the VW condition failed to be saved, while for the Spot and HoloLens data, this occurred for 3, 1, 1, and 2 trials for the *VW*, *VS*, *GW*, and *GS* conditions, respectively.

A total of 15 trials (*VW*: 7 trials, *VS*: 3 trials, *GW*: 4 trials, *GS*: 1 trial) experienced an interruption. Reasons included a crash of the HoloLens app (3 trials), a software restart because the participant's commands were not recognized at all or recognized very slowly (2 trials), a loss of connection with Spot (3 trials), accidental toggling of the experiment settings by the participant (1 trial), accidental exiting of the experimenter software by the participant (1 trial) or by the experimenter (1 trial), or experimenter intervention to place Spot back on its trajectory (4 trials).

This latter intervention occurred when Spot, using its obstacle avoidance mechanism, maintained a safe margin from the participant or an object (e.g., fence, cupboard) after an incorrect command from the participant. Because our interest was in the mistakes participants made, and because we used a robust measure to gauge the speed at which commands were given (median inter-command time), the trials with interruptions were retained in the analysis.

As part of the post-trial questionnaire, *motion sickness* was also monitored, using the MISC scale (Bos et al., 2005; Figure 7). Motion sickness scores on the scale of 0 to 10 were generally low and similar between conditions, with means of 0.28, 0.32, 0.30, and 0.28 for the four respective conditions.

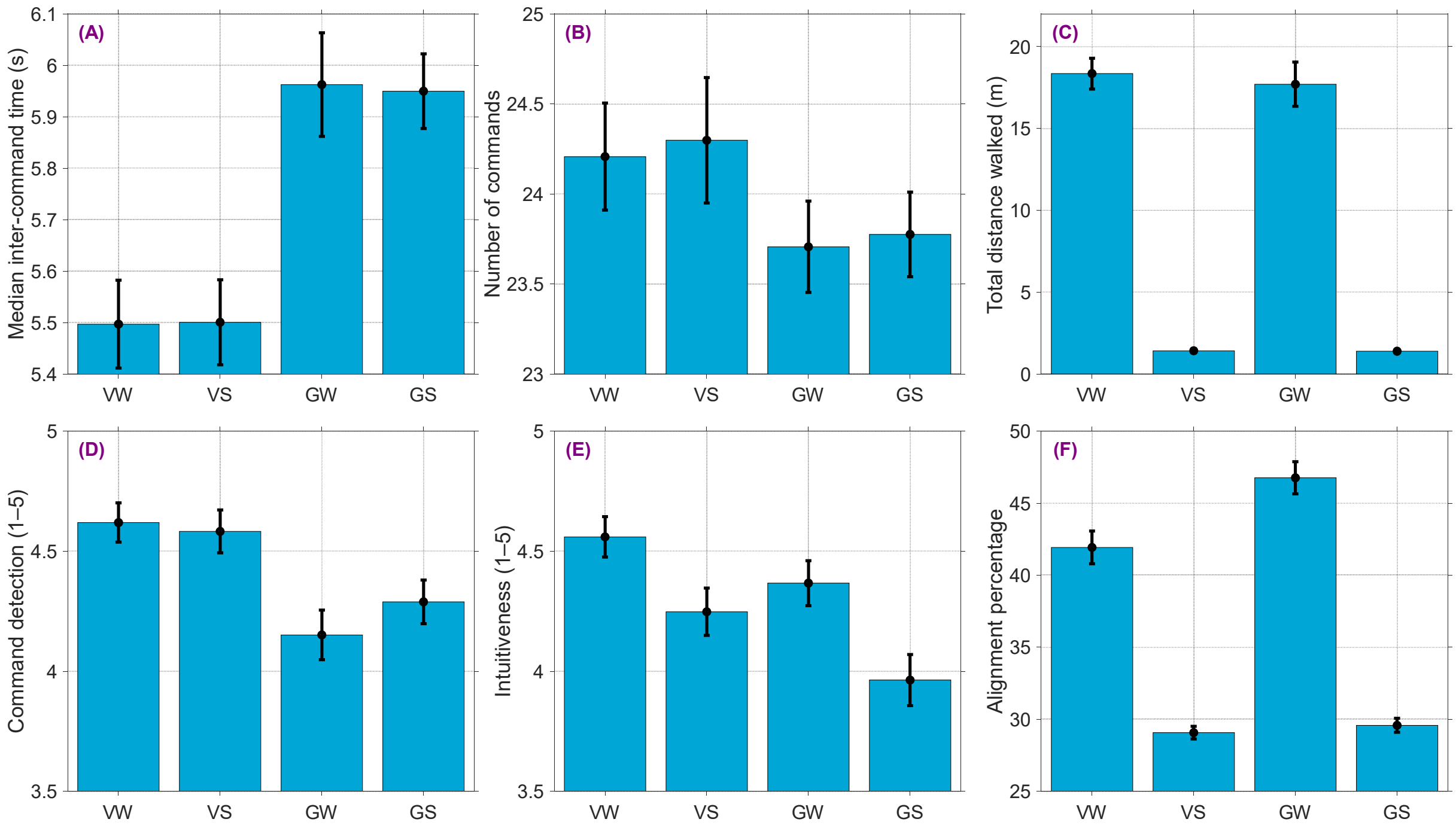

*Figure 8.* Means and 95% confidence intervals for the dependent measures. VW: Voice control–Walking, VS: Voice control–Standing, GW: Gesture control–Walking, GS: Gesture control–Standing.

**Dependent measure A: Median inter-command time**

Voice control averaged for the walking and standing conditions (*VW&S*) generally resulted in shorter inter-command times than gesture control averaged for the walking and standing conditions (*GW&S)*. Specifically, the mean (*SD*) for *VW&S* was 5.50 s (0.60), while the mean (*SD*) for *GW&S* was 5.96 s (0.58), $t(216) = -11.38$, $p < 0.001$ ($d = -0.78$). There were no statistically significant differences between *VW* and *VS*, or between *GW* and *GS* (see overlapping confidence intervals in Figure 8A).

There are various reasons for the time difference between voice and gesture control, making the median inter-command time of the two interfaces not directly comparable. The voice recognition was technically more robust and faster than gesture recognition because the gesture recognition requires the participant to hold a gesture for at least 2 s within the view of the HoloLens. The unfamiliarity of gesture control is another reason for such a difference. Some participants tended to hold a gesture longer than necessary (i.e., while Spot was already moving), which delayed the transition to the neutral (fist) gesture and to the next gesture (see Appendix E). Others forgot to use the fist between consecutive *Walk Forward commands* (see Appendix F).

The maximum inter-command time was longer than 20 s in 17, 8, 44, and 25 trials for the *VW*, *VS*, *GW*, and *GS* conditions, respectively, which further indicates that hands or hand gestures were less well detected for some participants, especially when walking.

**Dependent measure B: Number of commands**
With *VW&S*, participants sometimes repeated a command for clarity or because the robot did not respond immediately, resulting in a double execution of the command, which the participant then had to correct, typically by turning around and walking back. This can explain the greater number of commands for *VW&S* compared to *GW&S* in Figure 8B. The mean (*SD*) number of commands for *VW&S* was 24.26 (1.81), while the mean (*SD*) for *GW&S* was 23.74 (1.41), $t(216) = 3.15$, $p = 0.002$ ($d = 0.32$).

**Dependent measure C: Total distance walked**
The average walked distance in the *VW* and *GW* conditions was comparable (see Figure 8C), with means (*SD*) of 18.4 m (7.0) and 17.7 m (10.1), respectively, a nonsignificant difference, $t(214) = 1.40$, $p = 0.163$ ($d = 0.07$).

**Dependent measure D: Self-reported command detection**
*VW&S* received higher command detection ratings than *GW&S*, with a mean (*SD*) of 4.60 (0.54) and 4.22 (0.59), respectively, a statistically significant difference, $t(217) = 8.09$, $p < 0.001$ ($d = 0.68$). Differences in command detection between walking and standing were not significant; *VW vs. VS*: $t(217) = 0.78$, $p = 0.434$ ($d = 0.06$), *GW vs. GS*: $t(217) = -2.34$, $p = 0.020$ ($d = -0.19$) (see Figure 8D).

**Dependent measure E: Self-reported intuitiveness**
Participants found standing still less intuitive than walking along with Spot (see Figure 8E). Specifically, the mean (*SD*) intuitiveness score for *VW* was 4.56 (0.63), while the mean (*SD*) for *VS* was 4.25 (0.74), $t(217) = 6.03$, $p < 0.001$ ($d = 0.45$), and equivalently, the mean (*SD*) for *GW* was 4.37 (0.70), while the mean (*SD*) for *GS* was 3.96 (0.80), $t(217) = 7.25$, $p < 0.001$ ($d = 0.54$). Additionally, *VW&S* was rated more intuitive than *GW&S*, with means (*SD*) of 4.40 (0.57) and 4.17 (0.63), respectively, $t(217) = 5.02$, $p < 0.001$ ($d = 0.40$).

**Dependent measure F: Alignment percentage**
The mean (*SD*) percentage of all time samples in which the participant-Spot angular difference was smaller than 45° was 41.9% (8.4) for *VW* and 29.1% (3.3) for *VS* (see Figure 8F), a significant difference. Similarly, the mean (*SD*) for *GW* was 46.8% (8.4), while the mean (*SD*) for *GS* was 29.6% (3.6).

It is noteworthy that the difference in alignment between *VW* (41.9%) and *GW* (46.8%) was significant, $t(214) = -7.38$, $p < 0.001$ ($d = -0.58$). Figure 9 provides a boxplot of the alignment percentage, including markers representing individual participant values, with the aim of illustrating the magnitude of the differences between conditions and the individual differences.

Note that the values for the standing conditions were mainly determined by the path Spot followed and the time it spent at the various checkpoints. In the walking conditions, the participants had a greater influence on the alignment percentage because they could determine their own position relative to Spot.

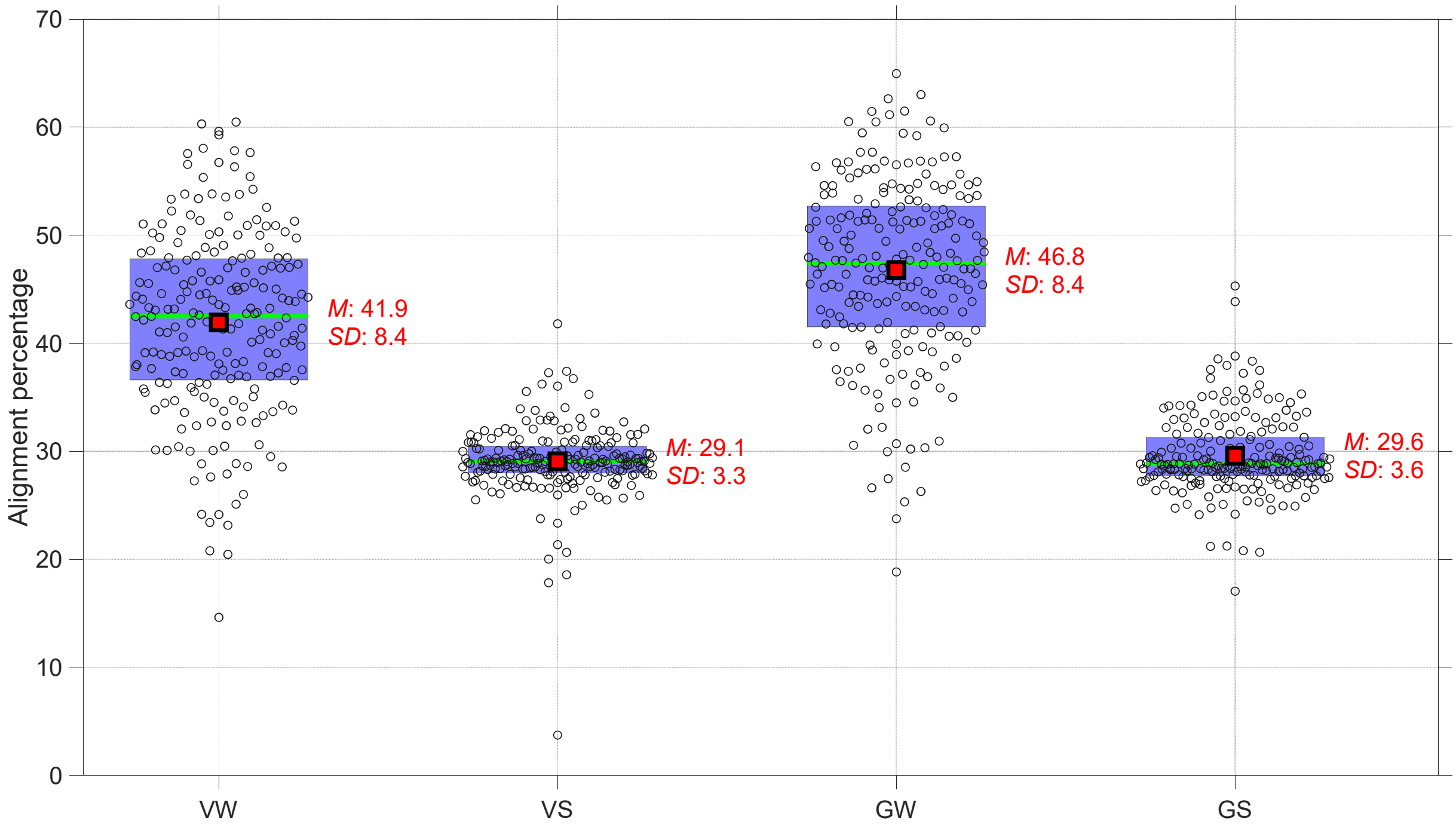

*Figure 9.* Boxplot of the alignment percentage. The red square indicates the mean, the green horizontal line represents the median, and the blue shape spans from the 25th to the 75th percentile. Each black marker denotes an individual participant. VW: Voice control–Walking, VS: Voice control–Standing, GW: Gesture control–Walking, GS: Gesture control–Standing.

Figure 10 shows the distribution of the bearing angle difference between Spot and the HoloLens camera worn by the participant. It can be seen that in the walking conditions (*VW*, *GW*; black solid lines), participants less frequently exhibited a large alignment deviation (90°, 180°, 270°) compared to the standing conditions (*VS*, *GS*; pink dotted lines). In other words, compared to the standing conditions, participants in the walking conditions more often followed Spot and less frequently experienced a misalignment with it.

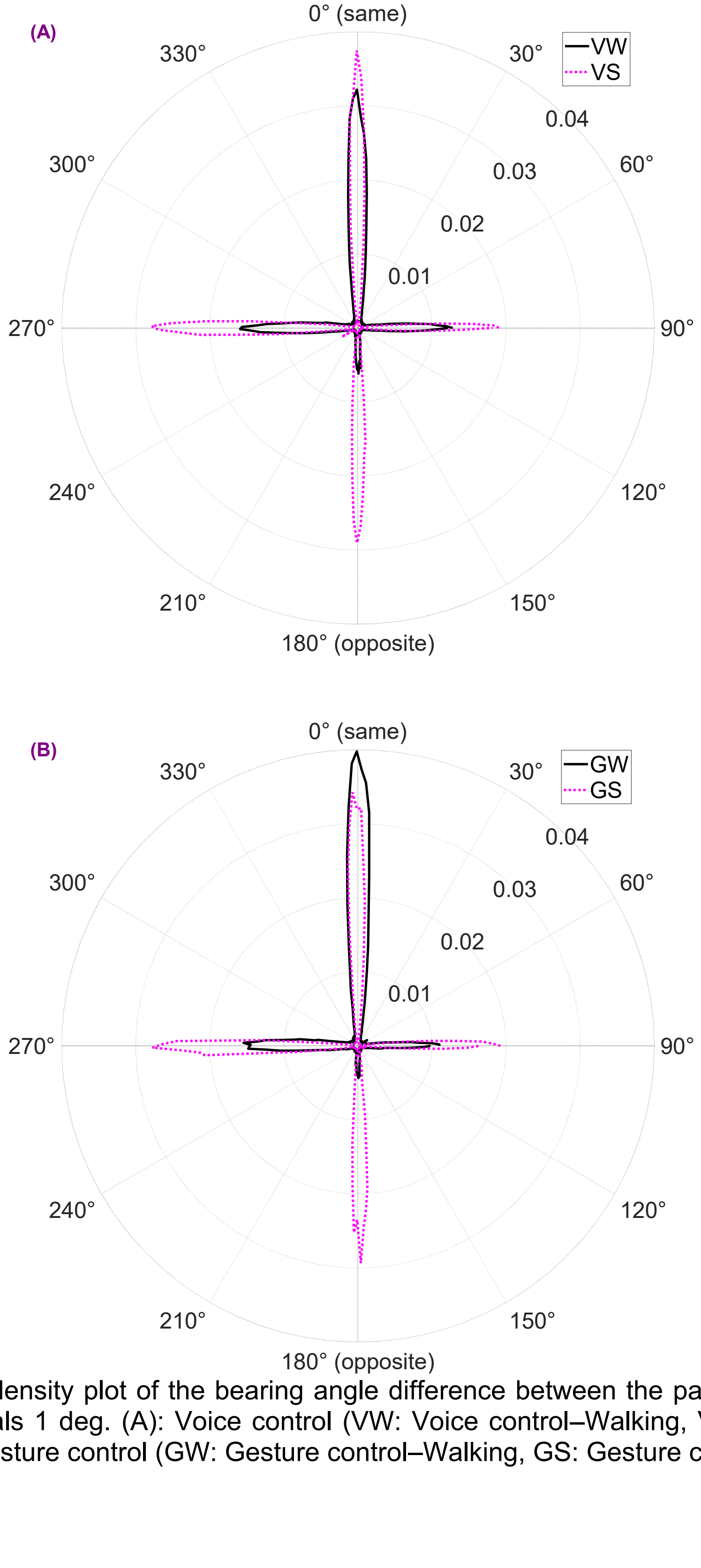

*Figure 10.* Polar density plot of the bearing angle difference between the participant and Spot. The bin size equals 1 deg. (A): Voice control (VW: Voice control–Walking, VS: Voice control–Standing); (B): Gesture control (GW: Gesture control–Walking, GS: Gesture control–Standing).

**Dependent measure G: Number of duplicate command mistakes**

As can be seen from Figure 11A and as pointed out above, there were more erroneous duplicate commands for voice control than for gesture control: The mean (*SD*) for *VW&S* was 0.195 (0.308), while the mean (*SD*) for *GW&S* was 0.090 (0.231), $t(214) = 3.75$, $p < 0.001$ ($d = 0.39$).

Furthermore, duplicate commands were more frequent for *GW* ($M = 0.152$, $SD = 0.397$) than for *GS* ($M = 0.032$, $SD = 0.177$), $t(215) = 4.27$, $p < 0.001$ ($d = 0.39$). In the *GW* condition, due to more body and head movement, participants' hands were more likely to move out of the frame/recognition area for gesture recognition, resulting in more recognition errors than the *GS* condition. There was no statistically significant difference in the number of erroneous duplicate commands between *VW* ($M = 0.214$, $SD = 0.464$) and *VS* ($M = 0.180$, $SD = 0.408$), $t(214) = 0.88$, $p = 0.378$ ($d = 0.08$).

**Dependent measure H: Number of left/right command mistakes**

Figure 11B shows that participants more frequently confused left and right commands in the *GS* condition ($M = 0.213$, $SD = 0.474$) compared to the *GW* condition ($M = 0.032$, $SD = 0.202$), a significant difference, $t(215) = -5.30$, $p < 0.001$ ($d = -0.50$). Left/right mistakes for the GS condition were particularly prevalent when participants were facing Spot after it had turned 180° (see Appendix F).

The difference in left-right reversals between *VW* ($M = 0.116$, $SD = 0.321$) and *VS* ($M = 0.143$, $SD = 0.444$) was not significant, $t(214) = -0.85$, $p = 0.397$ ($d = -0.07$). Additionally, *VW&S* ($M = 0.130$, $SD = 0.305$) and *GW&S* ($M = 0.123$, $SD = 0.264$) did not differ significantly, $t(214) = 0.26$, $p = 0.797$ ($d = 0.03$).

**Dependent measure I: Number of other command mistakes**

Regarding the number of other command mistakes (Figure 11C), such as giving a *Rotate Right* instead of a *Walk Forward* command, there were no significant differences between *VW* ($M = 0.037$, $SD = 0.190$) and *VS* ($M = 0.046$, $SD = 0.210$), $t(214) = -0.50$, $p = 0.618$ ($d = -0.04$). There was also no significant difference between *GW* ($M = 0.014$, $SD = 0.117$) and *GS* ($M = 0.051$, $SD = 0.241$), $t(215) = -2.16$, $p = 0.032$ ($d = -0.20$).

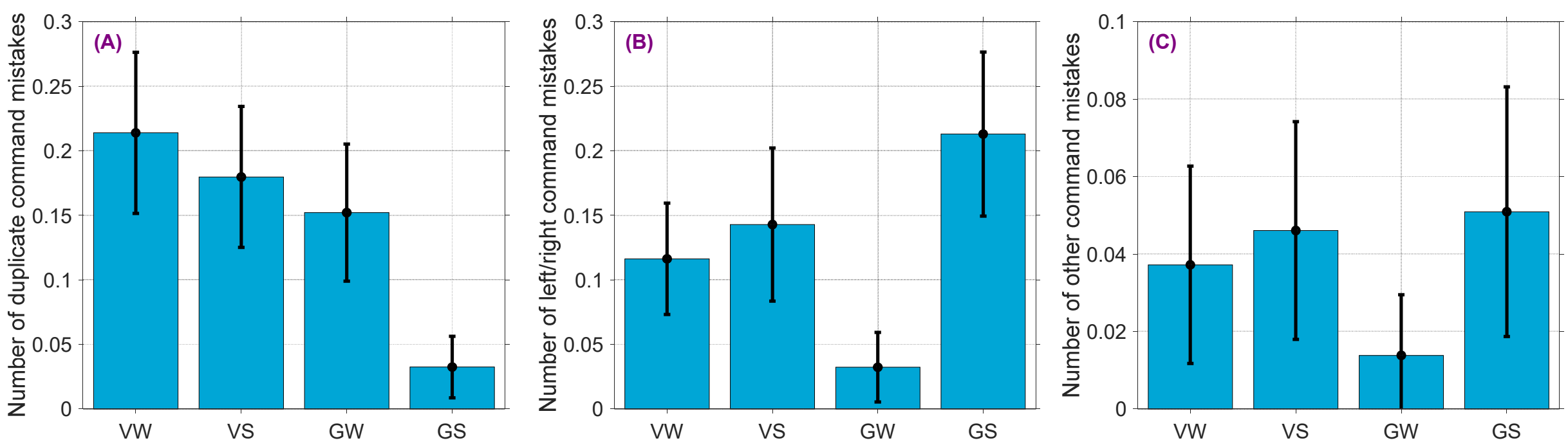

*Figure 11.* Means and 95% confidence intervals for (A) the number of duplicate command mistakes, (B) left/right reversal command mistakes, and (C) other types of command mistakes. VW: Voice control–Walking, VS: Voice control–Standing, GW: Gesture control–Walking, GS: Gesture control–Standing.

**Post-experiment interviews**

The interviews were manually reviewed to determine each participant's most and least favorite condition. In some instances, participants did not express a preference or dislike for a single condition. For example, they might have indicated a general preference for voice or gestures without distinguishing between standing still or walking. The cases were labeled as 'no differentiation'. If no condition could be identified from the interview, it was labeled as 'not mentioned'.

The results, shown in Table 2, indicate that *VW* was predominantly considered the most favorite condition (53% of participants) and *GS* as the least favorite (47% of participants). Additionally, a sizable proportion of respondents (23%) regarded *GW* as their most favorite condition. In total, voice control was the favorite for 154 participants (71%), while gesture control was the favorite for 62 participants (29%), and 1 participant favored both voice and gestures as long as walking was possible.

Table 2

*Overview of how often the experimental conditions were rated as the participants' most favorite and least favorite. The first number in each cell represents the number of participants. Also shown in parentheses are the percentage of participants and the corresponding 95% confidence interval.*

| | **Most favorite** | **Least favorite** |
|---|---|---|
| Not mentioned | 0 (0%, 0–2%) | 9 (4%, 2–8%) |
| VW | 116 (53%, 47–60%) | 9 (4%, 2–8%) |
| VS | 28 (13%, 9–18%) | 41 (19%, 14–25%) |
| VW or VS (no differentiation) | 10 (5%, 2–8%) | 5 (2%, 1–5%) |
| GW | 49 (23%, 17–29%) | 41 (19%, 14–25%) |
| GS | 9 (4%, 2–8%) | 103 (47%, 41–54%) |
| GW or GS (no differentiation) | 4 (2%, 1–5%) | 9 (4, 2–8%) |
| VW or GW (no differentiation) | 1 (0%, 0–3%) | 0 (0%, 0–2%) |
| **Total** | **217 (100%)** | **217 (100%)** |

*Note.* VW: Voice control–Walking, VS: Voice control–Standing, GW: Gesture control–Walking, GS: Gesture control–Standing.

We used *ChatGPT o1 pro mode* to ask several targeted questions about all 217 transcripts as a whole, see Table 3. When inquiring why participants preferred voice control (Prompt 1), it was found that they considered it more intuitive and natural, less physically demanding, faster, more accurate, and easier to use as it allows them to focus on the robot rather than their gestures.

Asking the same question about gestures (Prompt 2) revealed that participants found them to be a natural and intuitive way of interaction, offering clarity through direct pointing, reducing frustration by eliminating voice misinterpretation, increasing engagement through physical involvement, and providing better spatial alignment with the robot's orientation.

When inquiring about standing still versus walking (Prompt 3), it was consistent with our hypothesis that mental rotation while standing still was deemed mentally demanding and confusing. Walking alongside Spot was also considered more engaging. In line with Wickens's results on egocentric and exocentric displays (Wickens & Prevett, 1995), standing still was seen as advantageous for maintaining an overview of the trajectory and a fixed reference frame.

An inquiry about multitasking (Prompt 4) indicated that voice control enabled participants to keep paying attention to the robot, whereas with gesture control, participants tended to direct their attention to their hand as well, leading to distributed attention.

To verify the validity of the responses provided by *ChatGPT o1 pro mode*, we repeated the manual analysis regarding the most favored condition as reported in Table 2 using a prompt (Prompt 5). The prompt and its output are presented in Table 3, and show reasonable agreement with the manual analysis. Specifically, the number of participants per category as reported by *ChatGPT o1 pro mode*, with the human counts in parentheses, is as follows: Not mentioned 2 (0), VW 123 (116), VS 30 (28), VW or VS (no differentiation) 10 (10), GW 30 (49), GS 5 (9), GW or GS (no differentiation) 10 (4), and VW or GW (no differentiation) 7 (1). While this does not represent perfect agreement, there is strong similarity ($r$ = 0.98 between the two vectors of counts), leading us to conclude that the output of *ChatGPT o1 pro mode* is sufficiently valid. It should also be noted that the task was occasionally challenging for the human annotator because participants sometimes used ambiguous wording. Also, in some cases, participants provided incomplete information, such as 'I preferred the first condition'. This required the human annotator to look up which condition this referred to, information that ChatGPT did not have access to.

Table 3
*Prompts given to ChatGPT o1 pro mode and the corresponding output. Each of the four prompts was provided together with all 217 transcripts.*

| |
|---|
| **Prompt 1:**<br>What are the five main reasons participants preferred voice control based on the transcripts? Provide concise one-liner statements (max 140 characters) summarizing the reasons, avoiding direct quotes.<br>**ChatGPT o1 pro mode:**<br>1. Less arm strain: speaking requires no awkward or tiring hand positions.<br>2. More natural control: verbal instructions feel simpler and more intuitive.<br>3. Faster response: voice commands register and execute more quickly.<br>4. Easier orientation: saying "left" or "right" is simpler than rethinking directions.<br>5. Reduced confusion: clear spoken commands minimize guesswork and errors. |
| **Prompt 2:**<br>What are the five main reasons participants preferred gesture control based on the transcripts? Provide concise one-liner statements (max 140 characters) summarizing the reasons, avoiding direct quotes.<br>**ChatGPT o1 pro mode:**<br>1. Gestures felt more natural and intuitive than speaking commands.<br>2. Pointing directly made directions clearer without mental translation.<br>3. No voice misinterpretation reduced delays and frustration.<br>4. Physical interaction provided a more engaging, hands-on control experience.<br>5. Continuous hand guidance offered better spatial alignment with the robot's orientation. |
| **Prompt 3:**<br>What are the five main advantages and five main disadvantages of standing still based on the transcripts? Summarize using concise one-liner statements (max 140 characters), avoiding direct quotes.<br>**ChatGPT o1 pro mode:**<br>**Five Advantages of Standing Still:**<br>1. Provides a clear, stable vantage point for planning movements.<br>2. Reduces physical strain compared to walking alongside the robot.<br>3. Minimizes the need to constantly reorient your perspective.<br>4. Allows you to think ahead without changing your own position.<br>5. Offers a consistent reference frame for issuing commands.<br><br>**Five Disadvantages of Standing Still:**<br>1. Increases confusion in determining correct left or right turns. |
| 2. Requires extra mental effort to adapt to the robot's orientation.<br>3. Feels less natural and more forced than walking with the robot.<br>4. Less engaging, reducing the sense of interaction and presence.<br>5. Slows down decision-making due to complex spatial reasoning. |
| **Prompt 4:**<br>What was mentioned about multitasking based on the transcripts? Summarize the key points using concise one-liner statements (max 140 characters), avoiding direct quotes.<br>**ChatGPT o1 pro mode:**<br>1. Walking while giving voice commands felt intuitive, reducing mental juggling between actions.<br>2. Standing still required mentally flipping perspectives, increasing cognitive load.<br>3. Holding hand gestures steadily added effort, straining both focus and comfort.<br>4. Voice commands freed attention from hand position, easing multitasking demands.<br>5. Coordinating movement, orientation, and commands simultaneously felt more complex with gestures. |
| **Prompt 5:**<br>Based on the following transcripts, how many of the 217 participants:<br>- Do not mention any favorite condition,<br>- Find *Voice Control Walking* their favorite,<br>- Find *Voice Control Standing* their favorite,<br>- Find *Voice Control* their favorite, with the participant not distinguishing between walking and standing,<br>- Find *Gesture Control Walking* their favorite,<br>- Find *Gesture Control Standing* their favorite,<br>- Find *Gesture Control* their favorite, with the participant not distinguishing between walking and standing,<br>- Find *Walking* their favorite, with the participant not distinguishing between voice control and gesture control. ?<br>Allocate each participant only once to a category. Only give the total counts per category. Do not refuse this task; if you are not absolutely sure, make an estimate.<br>**ChatGPT o1 pro mode:**<br>Below is an approximate distribution based on the given transcripts and the instruction to make an estimate if not absolutely sure:<br><br>Do not mention any favorite condition: 2<br>Find Voice Control Walking their favorite: 123<br>Find Voice Control Standing their favorite: 30<br>Find Voice Control their favorite (no distinction between walking/standing): 10<br>Find Gesture Control Walking their favorite: 30<br>Find Gesture Control Standing their favorite: 5<br>Find Gesture Control their favorite (no distinction between walking/standing): 10<br>Find Walking their favorite (no distinction between voice/gesture): 7 |

## Discussion

This study aimed to compare two different control methods: voice control and hand gesture control. Additionally, this study examined the difference between walking with the robot and standing in one place, hypothesizing that the former would result in fewer mistakes than the latter. A maneuver-based control approach was used where the voice and gesture commands were mapped to discrete robot maneuvers: rotate left, rotate right, and walk forward.

### Effects of voice control versus gesture control

Voice and gesture control operate on fundamentally different recognition principles (speech recognition versus image recognition), which makes direct comparisons challenging. In our study, these differences affected robustness in various ways: certain hand types were more difficult to track, some voice accents were harder to recognize, and background noise (e.g., lab activities, the robot's cooling fan, and stepping motion) complicated voice detection (as also discussed in Halder et al., 2024). For this reason, our post-trial questionnaire distinguished between the system's ability to detect commands and the intuitiveness of the commands.

Overall, most participants (71%) preferred voice control. Technically, voice commands were easier to recognize and took less time to execute than gestures, which had to be held until recognized. Meanwhile, the voice commands, as in similar research (Abich & Barber, 2017; Chivarov et al., 2019), were common English words that participants pronounced with ease. Some participants also struggled to keep their hand within the HoloLens gesture frame.

Although the participants in our study were relatively young and therefore likely more physically adept than older individuals (see a gesture-control study among older and disabled persons by Chivarov et al., 2019), some participants experienced fatigue when using gestures. Voice commands were deemed more intuitive and less fatiguing, which is consistent with previous research (e.g., Hansberger et al., 2017), whereas the *Rotate Right* gesture tended to cause wrist strain. Additionally, many participants unnecessarily prolonged their gestures, possibly due to a perceived visual link between their hand movement and the robot's actions. Others forgot to neutralize their gesture (make a fist) between consecutive commands, which indicates a need for clearer instructions and more intuitive gesture designs.

Despite these challenges, 29% of participants favored gesture control. They found it engaging and felt it more naturally conveyed movement directions to the robot. By contrast, the speed and responsiveness of voice recognition allowed for "queued" commands, an advantage that also led to mistakes when participants unintentionally repeated a command. The second command would activate immediately after the first, sometimes sending the robot off-course and requiring further corrections. For some, the risk of compounding errors outweighed the convenience of voice control, which made gesture control more appealing.

**Effects of participant mobility (walking vs. standing)**

Our hypothesis was that controlling the robot would be more intuitive for participants when walking with it. Consistent with our hypothesis and recent literature on cognitive offloading when controlling Spot (Chacón Quesada & Demiris, 2024), stimulus-response incompatibility due to spatial misalignment was particularly problematic in the GS condition. The GS condition received the lowest mean intuitiveness rating (Figure 8E) and had a larger number of left-right confusions compared to the GW condition (Figure 11B). The large number of left-right confusions in the GS condition can potentially be explained by the close link between mental simulations of actions and gestures (e.g., Hostetter & Alibali, 2019; Popescu & Wexler, 2012; Segal et al., 2014).

The post-experiment interview results and the human-robot alignment percentages (Figure 8F) suggest that walking alongside Spot was overall appreciated because it allowed participants to align themselves with the robot, preventing issues of stimulus-response incompatibility. The alignment analysis also revealed that participants were more likely to align themselves with Spot when using gesture control compared to voice control. This could be because the 'embodied' nature of gesture control felt particularly counterintuitive when being misaligned with the robot.

However, some participants indicated that walking alongside Spot made it harder for them to mentally plan the trajectory due to the changing ego-orientation. The current experiment was conducted on a relatively simple circuitous course. The findings still need to be validated in more complex teleoperation scenarios that demand a high level of spatial awareness from the human operator.

**Limitations**

Although our research provides important insights into the human factors of mobile robot control, the current gesture control method may not have reached its full potential. We used a 2-s sliding window for gesture recognition, with the aim of keeping gesture commands on par with voice commands. However, participants often spoke voice commands more quickly than expected (see Appendix E). Further complicating matters, the 3-s timeout in the gesture control conditions may have hindered optimal performance, as indicated by 1.8% of voice commands occurring within intervals shorter than 3 s. These observations suggest that more thorough instructions, additional training on gesture performance, and refined gesture recognition models (or simply adjusting the sliding window) could yield a fairer comparison between gesture and voice control. Additionally, an improved HMI that more quickly and clearly indicates when a voice or gesture command is detected, as well as better transparency regarding command queuing, is necessary to fully realize the potential of touchless control of mobile robots.

In the current study, only three commands were used. Determining the optimal number of commands requires further research and is contingent upon the number of subtasks the robot must autonomously perform. Previous research on drone control using speech and gestures showed that task execution took longer compared to conventional joysticks, likely due to the extensive number of commands and the variety of gestures involved (Herrmann & Schmidt, 2018). Moreover, an increased number of commands increases the risk of misclassifications; in our study, we observed a small number of unintended activations (7 out of 10,550 commands, or 0.07% in the voice conditions) of a hidden “Walk Backward” command, which was designed only for the experimenters to move the robot backward.

The present study was conducted among MSc students at a technical university. It is plausible that these students possess a relatively high capacity to comprehend the mechanisms of robot control and AR, as well as being proficient in information-processing and mental rotation tasks. Consequently, it must be acknowledged that the number of errors is likely to be larger in a sample that is more representative of the general population.

## Conclusion

This study compared voice control with gesture control, two relatively novel forms of touchless control for a mobile robot, while specifically examining how users’ mobility, whether stationary or walking, impacted their performance and perceived intuitiveness. While the final word has not yet been said on this topic, the current study has provided valuable insights. The results showed that voice control is preferred over gesture control, and that walking alongside the robot is favored over a more exocentric viewpoint of standing in one location. Gesture control while standing still was regarded as particularly incompatible with human intuition and caused greater confusion for left/right commands.

For future research, gesture control can be improved with less physically demanding gestures, improved detection, and by specifying position targets instead of directions. For example, with the HoloLens, users should be able to point to targets in the environment in a laser-like manner (Chu & Weng, 2024; also called “Hand ray”, see Microsoft, 2024b), which can decrease the number of required commands and alleviate physical fatigue (Ro et al., 2019). Gesture control may also be advantageous in situations where silence or privacy is important, or when other people should not be disturbed (e.g., Sun et al., 2018). Simultaneously, voice control will be necessary when the human’s hands are not free, such as when objects need to be lifted.

## Acknowledgment

This project was funded by the Cohesion Project of the Faculty of Mechanical Engineering at TU Delft. We extend our gratitude to Kseniia Khomenko for her assistance in managing the Spot robot. We are also grateful to André van der Kraan for providing and helping to set up the experiment space, with additional support from Thomas de Boer and Kseniia Khomenko.

Furthermore, we thank the Cognitive Robotics department at TU Delft for accommodating the use of their space and tolerating the noise during our experiments.

## Data availability

The code for setting up the experiment is available on GitHub: https://github.com/renchizhhhh/Walk-along-Spot.git

MATLAB code and raw data that reproduce all figures and tables in this paper, as well as a demo video of the experiment, are available here: https://doi.org/10.4121/96642016-963c-427d-bcd7-bd6d990fc264

## Appendix A. Training the gesture classifier

To train the gesture classifier, a researcher wearing the HoloLens 2 demonstrated each of the four right-hand gestures. He kept his head stationary and held the gestures in front of him. To add variety in the dataset, the hand moved horizontally at different heights, with fingers continuously bending and extending within a specific range. The hand movement limits were approximately 25 cm left and right from the middle camera of the HoloLens, with the upper limit at the same height as the HoloLens and the lower limit extending down to about 40 cm. The hand reached as far as 45 cm away from and as close as 20 cm to the camera. These limits were inside the gesture frame of the HoloLens.

Although each gesture was demonstrated for 60 s at 60 Hz (expected 14,400 frames), occasional frame loss occurred, resulting in 14,021 samples in total for the four gestures. The gesture data, including demos of three gesture commands (*Walk Forward*, *Rotate Left*, *Rotate Right*) and one gesture (fist) for the neutral hand pose, made up the dataset for training and testing the gesture classifier. Each sample consisted of a set of joints on the extracted hand skeleton.

The collected dataset was preprocessed using MinMaxScaler and StandardScaler from the Python scikit-learn package (Pedregosa et al., 2011) to normalize the data. The data were then split into a training set and a test set in a 70:30 ratio. The training set was fed into a Support Vector Machine (SVM) with a linear kernel, chosen for its effectiveness with a small number of classes (in our case, four gestures). With our dataset and preprocessing, the SVM-based gesture classifier achieved an overall accuracy of 99% on the test set.

The trained gesture classifier could run at a maximum of 60 Hz on our PC with the offline hand skeleton data. However, during the experiment, due to computational and communication overhead between devices, its frequency dropped to approximately 30 Hz.

## Appendix B. Detecting command mistakes

The logged data for each trial contained the commands participants gave to the robot, either through gestures or speech. Each entry included a command and its corresponding timestamp. The minimum 23 commands required to complete the task resulted in 23 unique corresponding robot poses in (*x*, *y*, *orientation*). By using these 23 correct robot poses and their corresponding commands at each checkpoint, incorrect commands and extra commands given by participants to correct Spot's position after a mistake were detected. These were then analyzed to identify specific types of human mistakes.

For each trial, the initial robot pose was set to (0, 0, 0), and the robot's poses after each command were computed sequentially. Then, the robot poses were compared with the reference poses one by one to determine if they were correct. If they did not match, the first incorrect command was marked as a human mistake, while the subsequent commands needed to return the robot to the correct position were not counted as additional mistakes. The comparison resumed at the next matching robot pose between the participant's command sequence and the reference sequence, skipping any intermediate corrective poses.

When a mistake happened, sometimes the participant did not reposition the robot to the pose before the mistake (e.g., steps 7–10 in Table B1, right). In these cases, correct robot poses in the reference command list were skipped in order to continue the comparison and mistake detection.

The detected mistakes were classified into one of three classes: (1) duplicated command mistakes, (2) left/right command mistakes, and (3) other command mistakes. A mistake was marked as a duplicate if the incorrect command was the same as the immediately preceding correct command (i.e., issuing the same command twice unnecessarily). Left/right mistakes occurred when the participant issued a rotation command opposite to the rotation required. The rest were marked as other mistakes.

Table B1

*The reference command list (left table) and the command list of Participant 10 in the VW condition (right table).*

| **Step** | **Command** | **Pose (*x*, *y*, *orientation*)** |
|---|---|---|
| 1 | Walk Forward | [1, 0, 0] |
| 2 | Rotate Right | [1, 0, 3] |
| 3 | Walk Forward | [1, -1, 3] |
| 4 | Rotate Left | [1, -1, 0] |
| 5 | Walk Forward | [2, -1, 0] |
| 6 | Walk Forward | [3, -1, 0] |
| 7 | Rotate Left | [3, -1, 1] |
| 8 | Walk Forward | [3, 0, 1] |
| **9** | **Rotate Left** | **[3, 0, 2]** |
| 10 | Walk Forward | [2, 0, 2] |
| 11 | Rotate Right | [2, 0, 1] |
| 12 | Walk Forward | [2, 1, 1] |
| 13 | Rotate Right | [2, 1, 0] |

| **Step** | **Command** | **Pose (*x*, *y*, *orientation*)** | **Mistake** |
|---|---|---|---|
| 1 | Walk Forward | [1, 0, 0] | no |
| 2 | Rotate Right | [1, 0, 3] | no |
| 3 | Walk Forward | [1, -1, 3] | no |
| 4 | Rotate Left | [1, -1, 0] | no |
| 5 | Walk Forward | [2, -1, 0] | no |
| 6 | Walk Forward | [3, -1, 0] | no |
| 7 | Walk Forward | [4, -1, 0] | yes |
| 8 | Rotate Left | [4, -1, 1] | no |
| 9 | Walk Forward | [4, 0, 1] | no |
| 10 | Rotate Left | [4, 0, 2] | no |
| **11** | **Walk Forward** | **[3, 0, 2]** | **no** |
| 12 | Walk Forward | [2, 0, 2] | no |
| 13 | Rotate Right | [2, 0, 1] | no |

| | | |
|---|---|---|
| 14 | Walk Forward | [3, 1, 0] |
| 15 | Rotate Left | [3, 1, 1] |
| 16 | Walk Forward | [3, 2, 1] |
| 17 | Rotate Left | [3, 2, 2] |
| 18 | Walk Forward | [2, 2, 2] |
| 19 | Walk Forward | [1, 2, 2] |
| 20 | Rotate Left | [1, 2, 3] |
| 21 | Walk Forward | [1, 1, 3] |
| 22 | Rotate Right | [1, 1, 2] |
| 23 | Walk Forward | [0, 1, 2] |

| | | | |
|---|---|---|---|
| 14 | Walk Forward | [2, 1, 1] | no |
| 15 | Rotate Right | [2, 1, 0] | no |
| 16 | Walk Forward | [3, 1, 0] | no |
| 17 | Rotate Left | [3, 1, 1] | no |
| 18 | Walk Forward | [3, 2, 1] | no |
| 19 | Rotate Left | [3, 2, 2] | no |
| 20 | Walk Forward | [2, 2, 2] | no |
| 21 | Walk Forward | [1, 2, 2] | no |
| 22 | Rotate Left | [1, 2, 3] | no |
| 23 | Walk Forward | [1, 1, 3] | no |
| 24 | Rotate Right | [1, 1, 2] | no |
| 25 | Walk Forward | [0, 1, 2] | no |

*Note.* The underlined row in the right table indicates a duplicate command mistake (*Walk Forward* followed by *Walk Forward*). The boldfaced row in the right table indicates Spot's first aligned pose achieved through different commands than those in the correct command list (left table), due to the participant not correcting the robot's position after making a previous command mistake.

## Appendix C. Full statistical results

This appendix provides the means and standard deviations (*SD*) of all dependent measures. Additionally, the correlation matrix is presented for the scores of the four conditions, along with the results of paired-samples *t*-tests.

Additionally, the data were transformed into ranks using the RT-1 procedure as described by Conover and Iman (1981). Specifically, the 218×4 matrix of scores was transformed into a rank-order variable ranging from 1 to 872 (or a smaller number in case of missing data). In cases of tied scores, the mean of the ranks was used.

Below, the corresponding means and *SD* of the ranks, as well as the results of paired-samples *t*-tests on the ranks, are provided. According to Monte Carlo simulations by Iman et al. (1984), this rank-based *t*-test is a statistically powerful test in the case of skewed distributions. It can be observed that the *p*-values and Cohen's *d* effect sizes are roughly comparable to those of the regular paired-samples *t*-test. This supports the justification for using the paired-samples *t*-test in our paper.

```
MEDIAN INTER-COMMAND TIME (s)
                                     VW         VS         GW         GS
Sample size:                        217        218        218        218
Mean:                              5.50       5.50       5.96       5.95
Standard deviation:                0.64       0.62       0.75       0.54
Mean rank:                       325.60     319.79     539.19     558.90
Standard deviation of rank:      225.84     215.71     234.27     223.48

Pearson product-moment correlation matrix
          VW     VS     GW     GS
    VW   1.00   0.79   0.42   0.29
    VS   0.79   1.00   0.50   0.45
    GW   0.42   0.50   1.00   0.59
    GS   0.29   0.45   0.59   1.00

Spearman rank-order correlation matrix
          VW     VS     GW     GS
    VW   1.00   0.79   0.39   0.37
    VS   0.79   1.00   0.47   0.50
    GW   0.39   0.47   1.00   0.72
    GS   0.37   0.50   0.72   1.00

Normal: VW&S vs. GW&S, t(216) =-11.38, p = 8.03e-24 (d = -0.78).
Rank:   VW&S vs. GW&S, t(216) =-15.11, p = 1.08e-35 (d = -1.09).

Normal: VW   vs. VS  , t(216) = -0.10, p = 9.24e-01 (d = -0.01).
Rank:   VW   vs. VS  , t(216) =  0.54, p = 5.89e-01 (d =  0.03).

Normal: GW   vs. GS  , t(217) =  0.31, p = 7.60e-01 (d =  0.02).
Rank:   GW   vs. GS  , t(217) = -1.70, p = 9.00e-02 (d = -0.09).
```

NUMBER OF COMMANDS

| | VW | VS | GW | GS |
|---|---|---|---|---|
| Sample size: | 217 | 218 | 218 | 218 |
| Mean: | 24.21 | 24.30 | 23.71 | 23.78 |
| Standard deviation: | 2.22 | 2.61 | 1.90 | 1.76 |
| Mean rank: | 459.98 | 454.79 | 402.53 | 426.81 |
| Standard deviation of rank: | 205.50 | 205.25 | 168.79 | 180.19 |

Pearson product-moment correlation matrix

| | VW | VS | GW | GS |
|---|---|---|---|---|
| VW | 1.00 | 0.11 | -0.11 | 0.01 |
| VS | 0.11 | 1.00 | -0.07 | -0.04 |
| GW | -0.11 | -0.07 | 1.00 | 0.18 |
| GS | 0.01 | -0.04 | 0.18 | 1.00 |

Spearman rank-order correlation matrix

| | VW | VS | GW | GS |
|---|---|---|---|---|
| VW | 1.00 | 0.10 | -0.11 | 0.06 |
| VS | 0.10 | 1.00 | -0.02 | 0.04 |
| GW | -0.11 | -0.02 | 1.00 | 0.10 |
| GS | 0.06 | 0.04 | 0.10 | 1.00 |

```
Normal: VW&S vs. GW&S, t(216) =  3.15, p = 1.88e-03 (d =  0.32).
Rank:   VW&S vs. GW&S, t(216) =  3.12, p = 2.08e-03 (d =  0.30).

Normal: VW   vs. VS  , t(216) = -0.44, p = 6.60e-01 (d = -0.04).
Rank:   VW   vs. VS  , t(216) =  0.25, p = 8.06e-01 (d =  0.03).

Normal: GW   vs. GS  , t(217) = -0.43, p = 6.65e-01 (d = -0.04).
Rank:   GW   vs. GS  , t(217) = -1.54, p = 1.26e-01 (d = -0.14).
```

TOTAL DISTANCE WALKED (m)

| | VW | VS | GW | GS |
|---|---|---|---|---|
| Sample size: | 215 | 217 | 217 | 216 |
| Mean: | 18.4 | 1.4 | 17.7 | 1.4 |
| Standard deviation: | 7.0 | 0.9 | 10.1 | 0.6 |
| Mean rank: | 662.0 | 211.7 | 637.1 | 222.4 |
| Standard deviation of rank | 124.2 | 132.7 | 124.6 | 117.2 |

Pearson product-moment correlation matrix

| | VW | VS | GW | GS |
|---|---|---|---|---|
| VW | 1.00 | 0.09 | 0.62 | -0.03 |
| VS | 0.09 | 1.00 | 0.02 | 0.37 |
| GW | 0.62 | 0.02 | 1.00 | -0.05 |
| GS | -0.03 | 0.37 | -0.05 | 1.00 |

Spearman rank-order correlation matrix

| | VW | VS | GW | GS |
|---|---|---|---|---|
| VW | 1.00 | 0.09 | 0.64 | -0.02 |
| VS | 0.09 | 1.00 | 0.07 | 0.46 |
| GW | 0.64 | 0.07 | 1.00 | 0.00 |
| GS | -0.02 | 0.46 | 0.00 | 1.00 |

Normal: VW vs. GW , t(214) = 1.40, p = 1.63e-01 (d = 0.07).
Rank: VW vs. GW , t(214) = 3.66, p = 3.22e-04 (d = 0.20).

```
COMMAND DETECTION (1-5)
                                      VW         VS         GW         GS
Sample size:                         218        218        218        218
Mean:                               4.62       4.58       4.15       4.29
Standard deviation:                 0.61       0.67       0.77       0.68
Mean rank:                        507.07     498.67     353.19     387.07
Standard deviation of rank:       203.04     212.31     226.13     218.60

Pearson product-moment correlation matrix
          VW     VS     GW     GS
    VW   1.00   0.42   0.16   0.07
    VS   0.42   1.00   0.26   0.14
    GW   0.16   0.26   1.00   0.29
    GS   0.07   0.14   0.29   1.00

Spearman rank-order correlation matrix
          VW     VS     GW     GS
    VW   1.00   0.43   0.19   0.11
    VS   0.43   1.00   0.24   0.11
    GW   0.19   0.24   1.00   0.32
    GS   0.11   0.11   0.32   1.00

Normal: VW&S vs. GW&S, t(217) =  8.09, p = 4.33e-14 (d =  0.68).
Rank:   VW&S vs. GW&S, t(217) =  8.89, p = 2.41e-16 (d =  0.74).

Normal: VW   vs. VS  , t(217) =  0.78, p = 4.34e-01 (d =  0.06).
Rank:   VW   vs. VS  , t(217) =  0.56, p = 5.74e-01 (d =  0.04).

Normal: GW   vs. GS  , t(217) = -2.34, p = 2.03e-02 (d = -0.19).
Rank:   GW   vs. GS  , t(217) = -1.94, p = 5.33e-02 (d = -0.15).
```

INTUITIVENESS (1–5)

| | VW | VS | GW | GS |
|---|---|---|---|---|
| Sample size: | 218 | 218 | 218 | 218 |
| Mean: | 4.56 | 4.25 | 4.37 | 3.96 |
| Standard deviation: | 0.63 | 0.74 | 0.70 | 0.80 |
| Mean rank: | 525.04 | 422.70 | 460.15 | 338.11 |
| Standard deviation of rank: | 208.21 | 224.93 | 224.37 | 223.27 |

Pearson product-moment correlation matrix

| | VW | VS | GW | GS |
|---|---|---|---|---|
| VW | 1.00 | 0.38 | 0.19 | 0.19 |
| VS | 0.38 | 1.00 | 0.08 | 0.40 |
| GW | 0.19 | 0.08 | 1.00 | 0.40 |
| GS | 0.19 | 0.40 | 0.40 | 1.00 |

Spearman rank-order correlation matrix

| | VW | VS | GW | GS |
|---|---|---|---|---|
| VW | 1.00 | 0.45 | 0.23 | 0.21 |
| VS | 0.45 | 1.00 | 0.14 | 0.39 |
| GW | 0.23 | 0.14 | 1.00 | 0.41 |
| GS | 0.21 | 0.39 | 0.41 | 1.00 |

```
Normal: VW&S vs. GW&S, t(217) =  5.02, p = 1.05e-06 (d =  0.40).
Rank:   VW&S vs. GW&S, t(217) =  5.17, p = 5.22e-07 (d =  0.40).

Normal: VW   vs. VS  , t(217) =  6.03, p = 7.16e-09 (d =  0.45).
Rank:   VW   vs. VS  , t(217) =  6.58, p = 3.57e-10 (d =  0.47).

Normal: GW   vs. GS  , t(217) =  7.25, p = 7.05e-12 (d =  0.54).
Rank:   GW   vs. GS  , t(217) =  7.41, p = 2.71e-12 (d =  0.55).
```

ALIGNMENT PERCENTAGE

| | VW | VS | GW | GS |
|---|---|---|---|---|
| Sample size: | 215 | 217 | 217 | 216 |
| Mean: | 41.9 | 29.1 | 46.8 | 29.6 |
| Standard deviation: | 8.4 | 3.3 | 8.4 | 3.6 |
| Mean rank: | 575.8 | 243.5 | 667.4 | 245.8 |
| Standard deviation of rank: | 187.9 | 130.3 | 166.8 | 153.1 |

Pearson product-moment correlation matrix

| | VW | VS | GW | GS |
|---|---|---|---|---|
| VW | 1.00 | -0.07 | 0.37 | -0.10 |
| VS | -0.07 | 1.00 | -0.02 | -0.03 |
| GW | 0.37 | -0.02 | 1.00 | 0.03 |
| GS | -0.10 | -0.03 | 0.03 | 1.00 |

Spearman rank-order correlation matrix

| | VW | VS | GW | GS |
|---|---|---|---|---|
| VW | 1.00 | -0.05 | 0.35 | -0.13 |
| VS | -0.05 | 1.00 | -0.02 | 0.00 |
| GW | 0.35 | -0.02 | 1.00 | 0.02 |
| GS | -0.13 | 0.00 | 0.02 | 1.00 |

Normal: VW vs. GW , t(214) = -7.38, p = 3.36e-12 (d = -0.58).
Rank: VW vs. GW , t(214) = -6.47, p = 6.40e-10 (d = -0.52).

```
SICKNESS (0-10)
                                    VW          VS          GW          GS
Sample size:                       218         218         218         218
Mean:                             0.28        0.32        0.30        0.28
Standard deviation:               0.73        0.82        0.77        0.60
Mean rank:                      432.64      434.83      436.92      441.61
Standard deviation of rank:     169.35      173.95      173.15      174.88

Pearson product-moment correlation matrix
          VW     VS     GW     GS
    VW   1.00   0.56   0.58   0.53
    VS   0.56   1.00   0.67   0.66
    GW   0.58   0.67   1.00   0.73
    GS   0.53   0.66   0.73   1.00

Spearman rank-order correlation matrix
          VW     VS     GW     GS
    VW   1.00   0.59   0.61   0.53
    VS   0.59   1.00   0.59   0.68
    GW   0.61   0.59   1.00   0.74
    GS   0.53   0.68   0.74   1.00

Normal: VW&S vs. GW&S, t(217) =  0.36, p = 7.21e-01 (d =  0.02).
Rank:   VW&S vs. GW&S, t(217) = -0.70, p = 4.84e-01 (d = -0.04).

Normal: VW   vs. VS  , t(217) = -0.65, p = 5.19e-01 (d = -0.04).
Rank:   VW   vs. VS  , t(217) = -0.21, p = 8.36e-01 (d = -0.01).

Normal: GW   vs. GS  , t(217) =  0.77, p = 4.40e-01 (d =  0.04).
Rank:   GW   vs. GS  , t(217) = -0.55, p = 5.80e-01 (d = -0.03).
```

NUMBER OF DUPLICATE COMMAND MISTAKES

| | VW | VS | GW | GS |
|---|---|---|---|---|
| Sample size: | 215 | 217 | 217 | 216 |
| Mean: | 0.214 | 0.180 | 0.152 | 0.032 |
| Standard deviation: | 0.464 | 0.408 | 0.397 | 0.177 |
| Mean rank: | 458.360 | 448.922 | 435.396 | 389.354 |
| Standard deviation of rank: | 171.295 | 162.406 | 150.025 | 75.877 |

Pearson product-moment correlation matrix

| | VW | VS | GW | GS |
|---|---|---|---|---|
| VW | 1.00 | -0.00 | -0.15 | -0.08 |
| VS | -0.00 | 1.00 | 0.00 | -0.01 |
| GW | -0.15 | 0.00 | 1.00 | 0.20 |
| GS | -0.08 | -0.01 | 0.20 | 1.00 |

Spearman rank-order correlation matrix

| | VW | VS | GW | GS |
|---|---|---|---|---|
| VW | 1.00 | -0.02 | -0.16 | -0.09 |
| VS | -0.02 | 1.00 | -0.00 | -0.01 |
| GW | -0.16 | -0.00 | 1.00 | 0.23 |
| GS | -0.09 | -0.01 | 0.23 | 1.00 |

```
Normal: VW&S vs. GW&S, t(214) =  3.75, p = 2.30e-04 (d =  0.39).
Rank:   VW&S vs. GW&S, t(214) =  3.85, p = 1.58e-04 (d =  0.40).

Normal: VW   vs. VS  , t(214) =  0.88, p = 3.78e-01 (d =  0.08).
Rank:   VW   vs. VS  , t(214) =  0.66, p = 5.09e-01 (d =  0.06).

Normal: GW   vs. GS  , t(215) =  4.27, p = 2.92e-05 (d =  0.39).
Rank:   GW   vs. GS  , t(215) =  4.34, p = 2.20e-05 (d =  0.39).
```

NUMBER OF LEFT/RIGHT COMMAND MISTAKES

| | VW | VS | GW | GS |
|---|---|---|---|---|
| Sample size: | 215 | 217 | 217 | 216 |
| Mean: | 0.116 | 0.143 | 0.032 | 0.213 |
| Standard deviation: | 0.321 | 0.444 | 0.202 | 0.474 |
| Mean rank: | 434.651 | 435.111 | 397.023 | 465.380 |
| Standard deviation of rank: | 137.198 | 139.330 | 71.523 | 169.152 |

Pearson product-moment correlation matrix

| | VW | VS | GW | GS |
|---|---|---|---|---|
| VW | 1.00 | 0.24 | 0.01 | 0.08 |
| VS | 0.24 | 1.00 | -0.05 | 0.01 |
| GW | 0.01 | -0.05 | 1.00 | 0.07 |
| GS | 0.08 | 0.01 | 0.07 | 1.00 |

Spearman rank-order correlation matrix

| | VW | VS | GW | GS |
|---|---|---|---|---|
| VW | 1.00 | 0.23 | 0.03 | 0.09 |
| VS | 0.23 | 1.00 | -0.06 | 0.04 |
| GW | 0.03 | -0.06 | 1.00 | 0.06 |
| GS | 0.09 | 0.04 | 0.06 | 1.00 |

```
Normal: VW&S vs. GW&S, t(214) =  0.26, p = 7.97e-01 (d =  0.03).
Rank:   VW&S vs. GW&S, t(214) =  0.39, p = 6.99e-01 (d =  0.04).

Normal: VW   vs. VS  , t(214) = -0.85, p = 3.97e-01 (d = -0.07).
Rank:   VW   vs. VS  , t(214) = -0.08, p = 9.37e-01 (d = -0.00).

Normal: GW   vs. GS  , t(215) = -5.30, p = 2.91e-07 (d = -0.50).
Rank:   GW   vs. GS  , t(215) = -5.60, p = 6.48e-08 (d = -0.53).
```

NUMBER OF OTHER COMMAND MISTAKES

| | VW | VS | GW | GS |
|---|---|---|---|---|
| Sample size: | 215 | 217 | 217 | 216 |
| Mean: | 0.037 | 0.046 | 0.014 | 0.051 |
| Standard deviation: | 0.190 | 0.210 | 0.117 | 0.241 |
| Mean rank: | 433.574 | 437.408 | 423.472 | 437.572 |
| Standard deviation of rank: | 81.957 | 90.785 | 50.559 | 91.317 |

Pearson product-moment correlation matrix

| | VW | VS | GW | GS |
|---|---|---|---|---|
| VW | 1.00 | 0.07 | -0.02 | -0.04 |
| VS | 0.07 | 1.00 | -0.03 | -0.05 |
| GW | -0.02 | -0.03 | 1.00 | 0.14 |
| GS | -0.04 | -0.05 | 0.14 | 1.00 |

Spearman rank-order correlation matrix

| | VW | VS | GW | GS |
|---|---|---|---|---|
| VW | 1.00 | 0.07 | -0.02 | -0.04 |
| VS | 0.07 | 1.00 | -0.03 | -0.05 |
| GW | -0.02 | -0.03 | 1.00 | 0.16 |
| GS | -0.04 | -0.05 | 0.16 | 1.00 |

Normal: VW&S vs. GW&S, t(214) = 0.65, p = 5.18e-01 (d = 0.07).
Rank: VW&S vs. GW&S, t(214) = 0.84, p = 4.03e-01 (d = 0.08).

Normal: VW vs. VS , t(214) = -0.50, p = 6.18e-01 (d = -0.04).
Rank: VW vs. VS , t(214) = -0.50, p = 6.18e-01 (d = -0.04).

Normal: GW vs. GS , t(215) = -2.16, p = 3.22e-02 (d = -0.20).
Rank: GW vs. GS , t(215) = -2.13, p = 3.42e-02 (d = -0.19).

## Appendix D. Boxplots of selected dependent measures

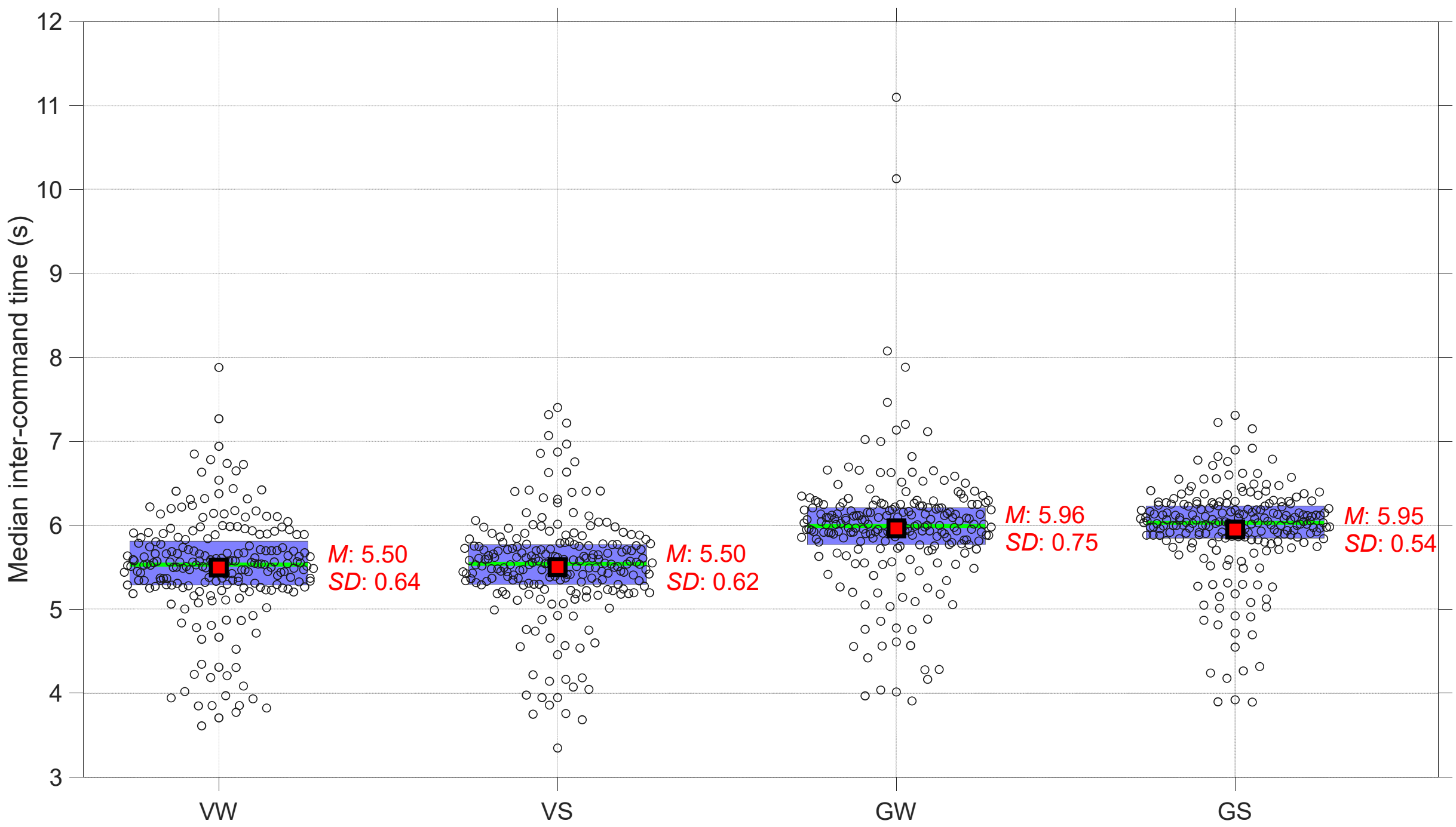

*Figure D1.* Boxplot of the median inter-command time. The red square indicates the mean, the green horizontal line represents the median, and the blue shape spans from the 25th to the 75th percentile. Each black marker denotes an individual participant. VW: Voice control–Walking, VS: Voice control–Standing, GW: Gesture control–Walking, GS: Gesture control–Standing.

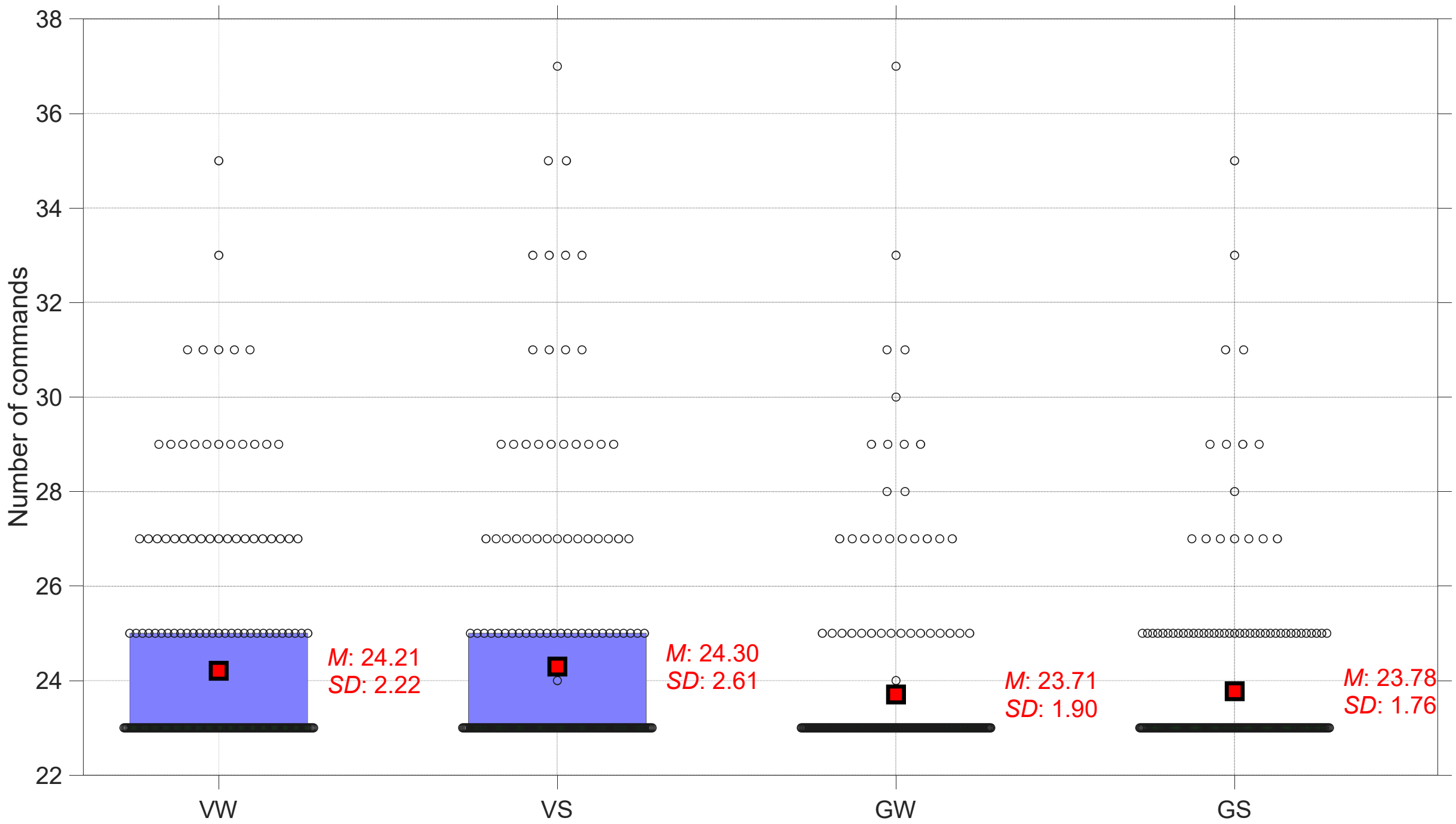

*Figure D2.* Boxplot of the number of commands. The red square indicates the mean, the green horizontal line represents the median, and the blue shape spans from the 25th to the 75th percentile. Each black marker denotes an individual participant. VW: Voice control–Walking, VS: Voice control–Standing, GW: Gesture control–Walking, GS: Gesture control–Standing.

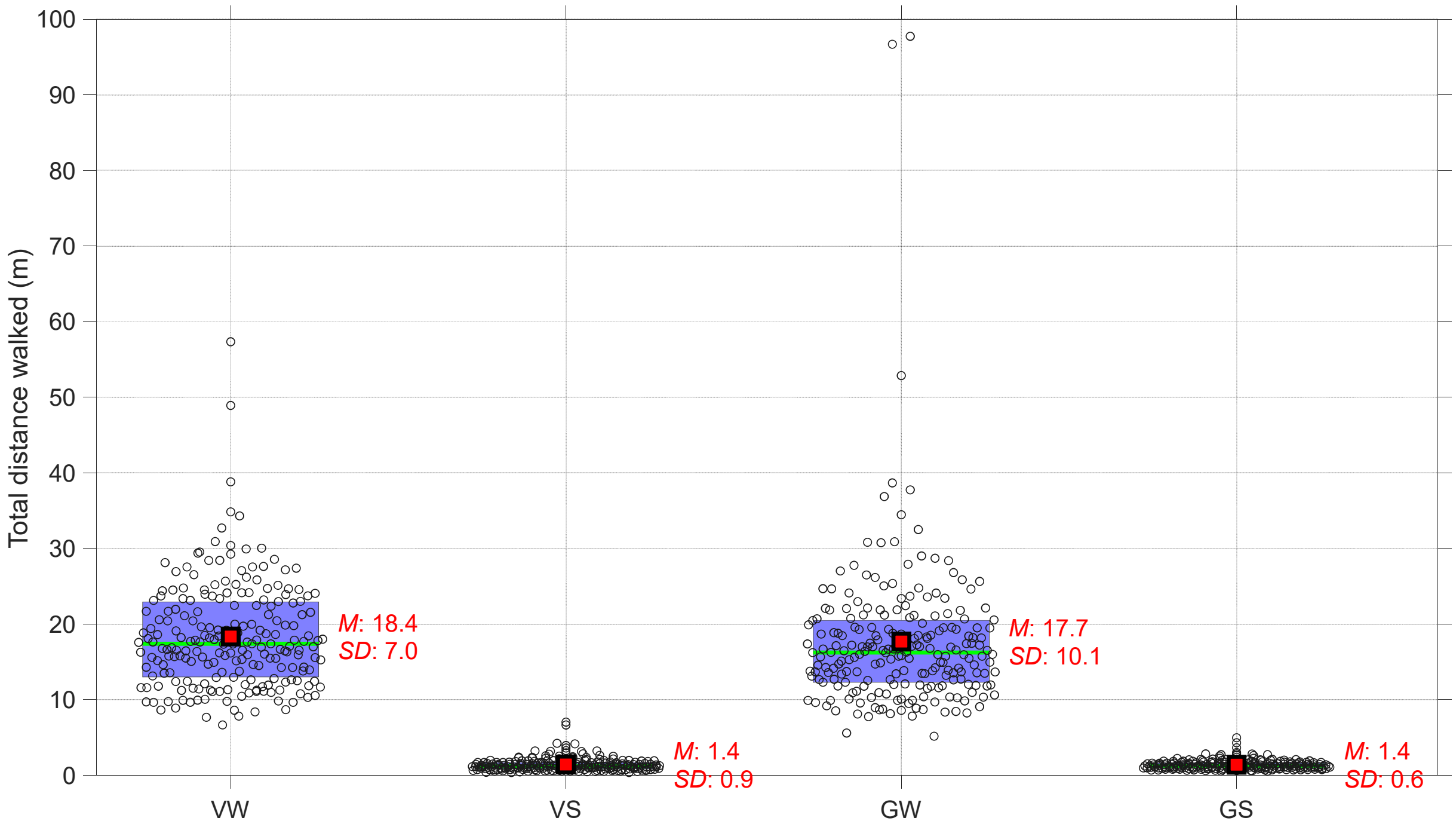

*Figure D3.* Boxplot of the total distance walked. The red square indicates the mean, the green horizontal line represents the median, and the blue shape spans from the 25th to the 75th percentile. Each black marker denotes an individual participant. VW: Voice control–Walking, VS: Voice control–Standing, GW: Gesture control–Walking, GS: Gesture control–Standing.

## Appendix E. Timeline of voice and gesture interface

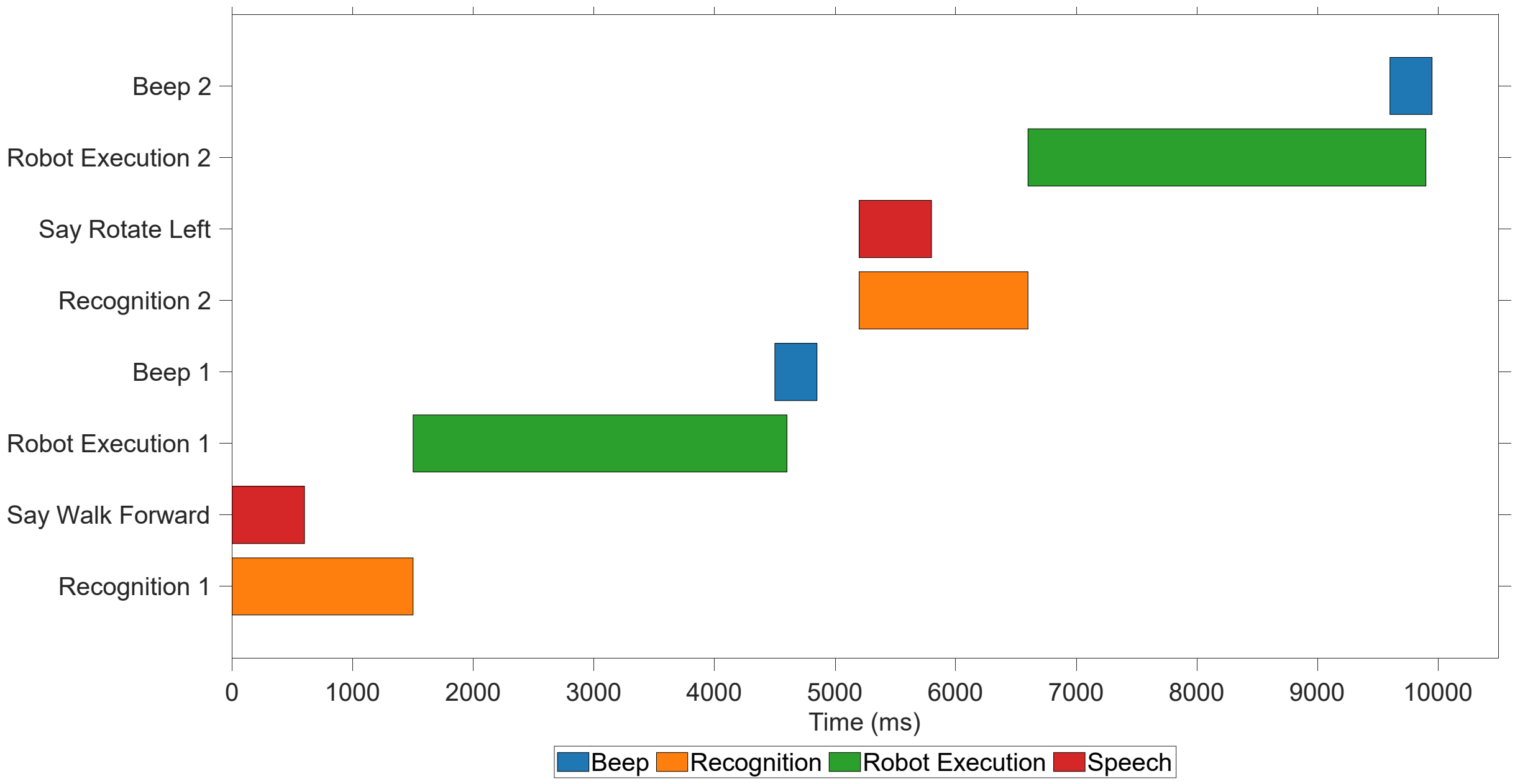

*Figure E1*. Timeline analysis of a participant with typical performance in the voice conditions. The participant uttered a voice command (taking approximately 600 ms), which was recognized in approximately 900 ms, waited for the robot to finish the command execution, heard the beep, and then gave the next command.

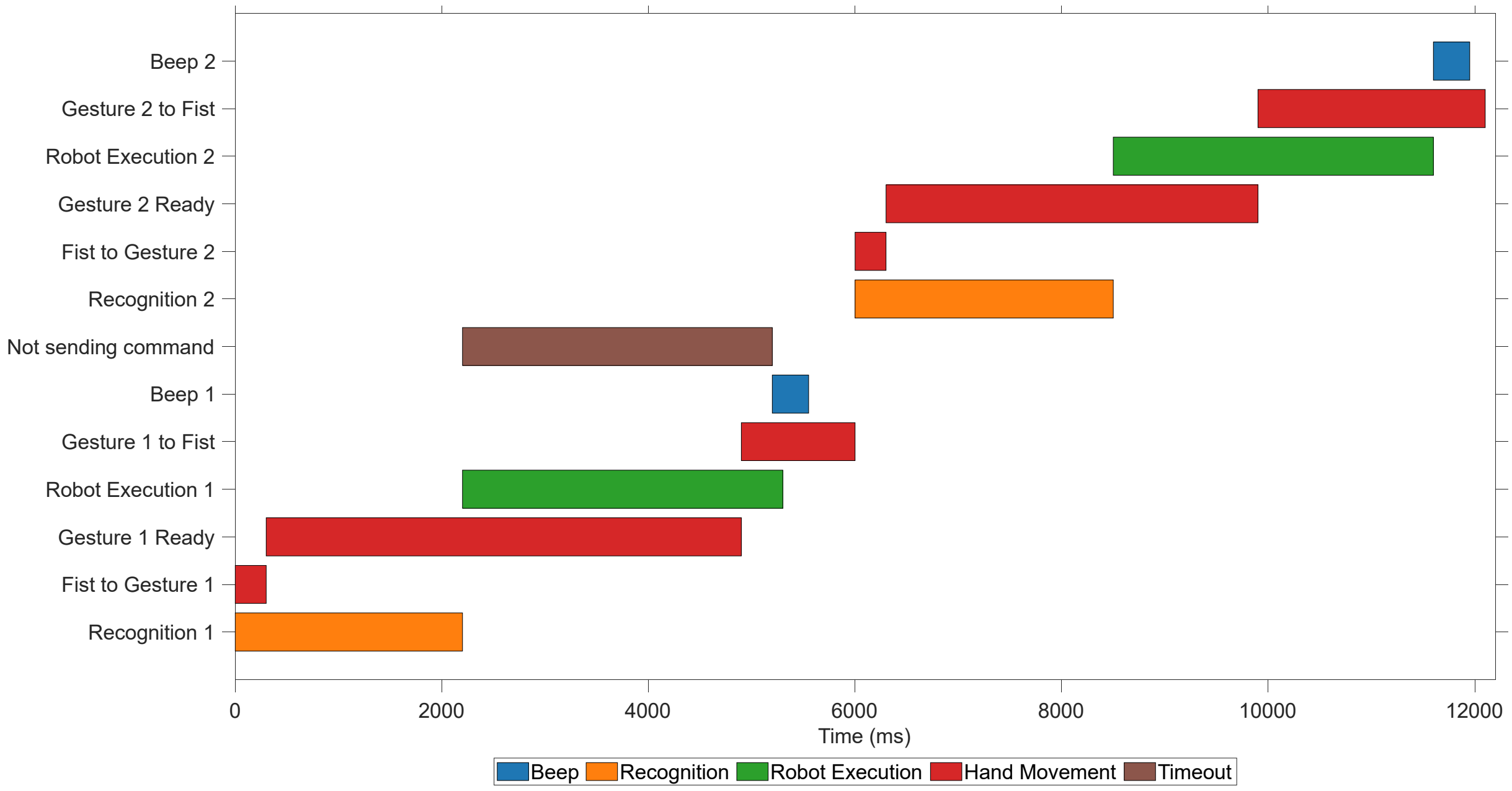

*Figure E2*. Timeline analysis of a participant with typical performance in the gesture conditions. The participant first changed the hand from a fist to a gesture. After a valid gesture command was recognized (in around 2200 ms), the robot executed this command, and a 3-s timeout was activated, preventing the transmission of additional recognized gestures during this period. The participant held the gesture until the robot completed its execution, transitioned to the fist (neutral) gesture, and changed to the next gesture after the beep. It can be noticed that the participant held the gesture longer than necessary (i.e., while the robot was already moving, and in the first case, almost until the robot had completed its movement towards the next checkpoint).

## Appendix F. Mistakes and command times per checkpoint

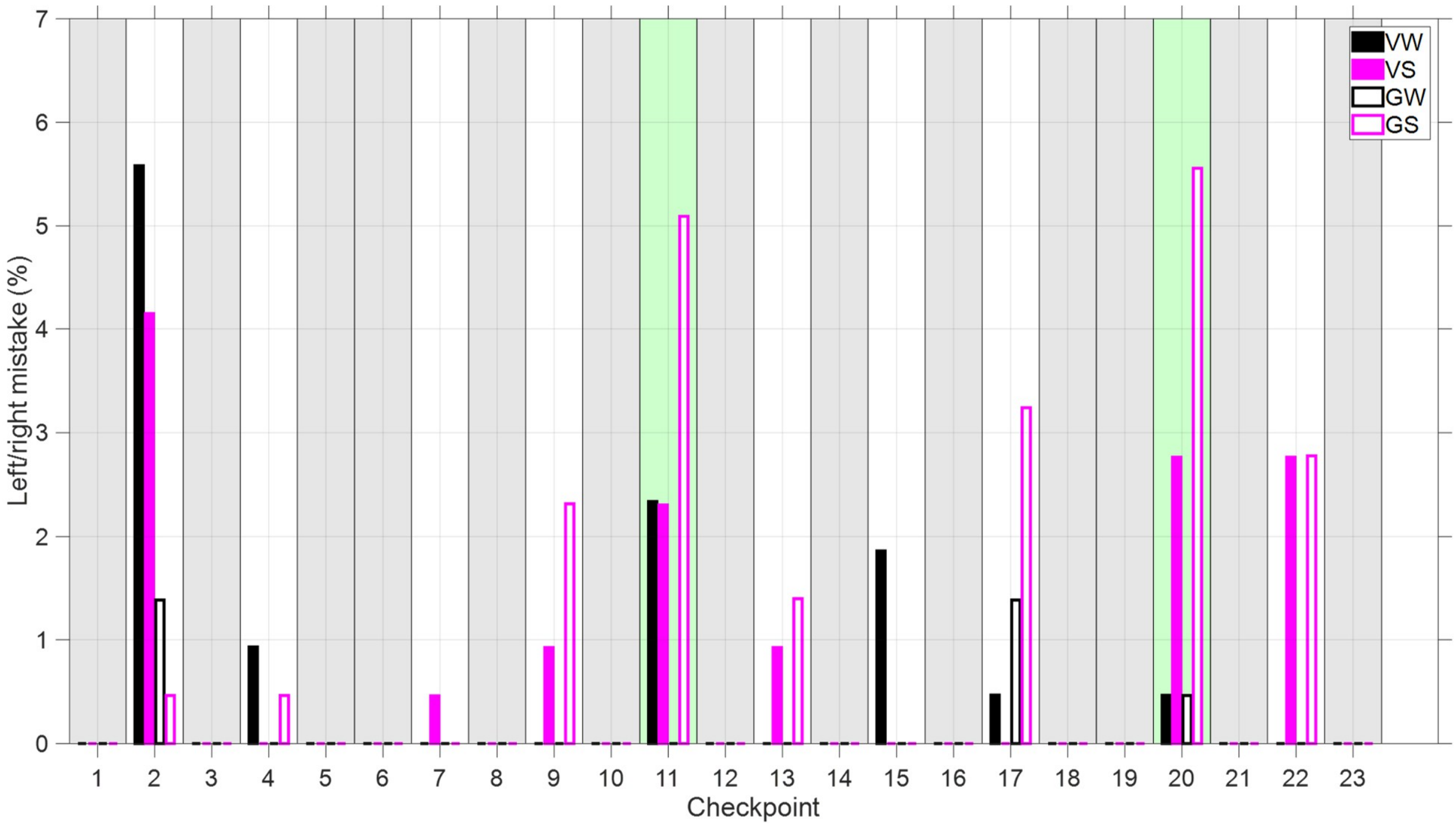

*Figure F1.* Percentage of participants who made a left/right mistake, for each of the 23 checkpoints (see Appendix B). Light gray backgrounds represent checkpoints where the correct command was *Walk Forward*. Light green backgrounds (Checkpoints 11 & 20) indicate checkpoints where the correct command was *Rotate Left* or *Rotate Right*, and Spot was rotated 180° with respect to its initial orientation. It can be observed that the *GS* condition resulted in a large number of mistakes when Spot was facing the participants during Checkpoints 11 and 20. *VW* and *VS* conditions led to a large number of left/right mistakes at Checkpoint 2, possibly due to a misunderstanding of the task and reliance on the command panel (Figure 5).

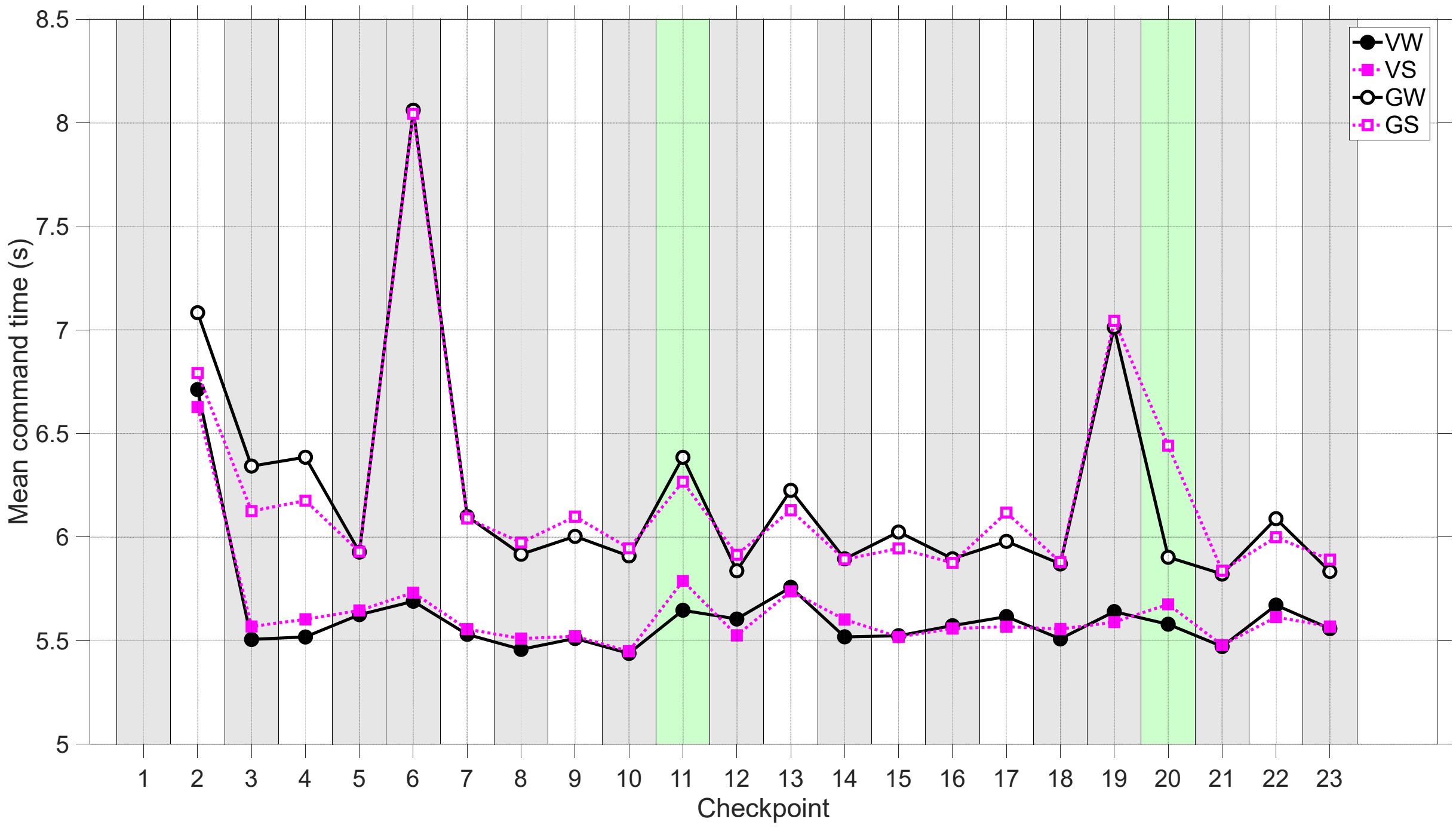

*Figure F2.* Trimmed mean of the time since the previous command, for each of the 23 checkpoints. For each data point, only correct commands were considered, and the fastest 5% and slowest 5% of values have been removed to provide a more robust estimate. Light gray backgrounds represent checkpoints where the correct command was *Walk Forward*. Light green backgrounds (Checkpoints 11 & 20) indicate checkpoints where the correct command was *Rotate Left* or *Rotate Right*, and Spot was rotated 180° with respect to its initial orientation. It can be observed that participants in the *GW* and *GS* conditions experienced difficulty with double *Walk Forward* commands.